\RequirePackage[l2tabu, orthodox]{nag}
\RequirePackage{snapshot}

\documentclass[9pt,twocolumn]{extarticle}

\makeatletter
\if@twocolumn
  \usepackage[dvips,letterpaper,top=0.5in, bottom=0.5in, left=0.75in, right=0.5in,includefoot,heightrounded]{geometry}
\else
  \usepackage[dvips,letterpaper,margin=1in,includefoot,heightrounded]{geometry}
\fi

\usepackage{srcltx}

\usepackage[russian,portuges,english]{babel}

\iflanguage{portuges}
    {\newcommand{\keywordname}{Palavras-chaves}}
    {\newcommand{\keywordname}{Keywords}}

\usepackage{amsmath}
\usepackage{amssymb,amsfonts}

\usepackage{abstract}

\usepackage{graphicx}
\usepackage[usenames,dvipsnames,svgnames,x11names]{xcolor}
\usepackage{subfigure}

\usepackage{booktabs}

\usepackage{setspace}
\usepackage{flushend}

\usepackage{cite}

\usepackage{hyperref}
\usepackage[normalem]{ulem}

\usepackage{enumerate}

\usepackage{algorithm}
\usepackage{algorithmic}

\usepackage{listings}

\usepackage[short,12hr]{datetime}
       \usepackage{fouriernc}

\newcommand{\PR}[1]{\ensuremath{\left[#1\right]}}
\newcommand{\PC}[1]{\ensuremath{\left(#1\right)}}

\makeatletter

\newenvironment{rsmallmatrix}{\null\,\vcenter\bgroup
  \Let@\restore@math@cr\default@tag
  \baselineskip6\ex@ \lineskip1.5\ex@ \lineskiplimit\lineskip
  \ialign\bgroup\hfil$\m@th\scriptstyle##$&&\thickspace\hfil
  $\m@th\scriptstyle##$\crcr
}{%
  \crcr\egroup\egroup\,%
}

\newcommand{\printtitle}{%
\makeatletter
\if@twocolumn

\twocolumn[%
  \maketitle
  \begin{onecolabstract}
    \myabstract
  \end{onecolabstract}
  \begin{center}
    \small
    \textbf{\keywordname}
    \\\medskip
    \mykeywords
  \end{center}
  \bigskip
]
\saythanks
\else
  \maketitle
  \begin{onecolabstract}
    \myabstract
  \end{onecolabstract}
  \begin{center}
    \small
    \textbf{\keywordname}
    \\\medskip
    \mykeywords
  \end{center}
  \bigskip
  \onehalfspacing
\fi
\makeatother
}

\author{%
A.~P.~Rad\"unz
\thanks{Programa de P\'os-Gradua\c c\~ao em Estat\'istica, Universidade Federal de Pernambuco, Recife, Brazil.
E-mail: \url{apr1@de.ufpe.br}}
\and
F.~M.~Bayer
\thanks{Departamento de Estat\'istica and LACESM, Universidade Federal de Santa Maria, Santa Maria, Brazil.
E-mail: \url{bayer@ufsm.br}}
\and
R.~J.~Cintra
\thanks{%
Signal Processing Group, CCEN,
UFPE, Brazil.
E-mail: \url{rjdsc@de.ufpe.br}}
}

\title{%
Low-complexity Rounded KLT Approximation for Image Compression}

\newcommand{\myabstract}{%
The Karhunen-Lo\`eve transform (KLT) is often used for data decorrelation and
dimensionality reduction.
Because
its computation depends on the matrix of covariances of the input signal, the use of the KLT in real-time applications is severely constrained by the difficulty in developing fast algorithms to implement it. In this context, this paper proposes a new class of low-complexity transforms that are obtained through the application of the round function to the elements of the KLT matrix.
The proposed transforms are evaluated considering figures of merit that measure the coding power and distance of the proposed approximations to the exact KLT and are also explored in image compression experiments.
Fast algorithms are introduced for the proposed approximate transforms.
It was shown that the proposed transforms perform well in image compression and require a low implementation cost.
}

\newcommand{\mykeywords}{%
Approximate KLT,
Image compression,
Karhunen-Lo\`eve transform,
Low-complexity transforms
}

\date{}

\date{\today\ @ \currenttime}

\begin{document}

\printtitle

\section{Introduction}
We are living in the era where
colossal amounts of data are generated every day, the Big Data era~\cite{goldston2008big}.
With increasing computational demands for analyzing large masses of data in real-time,
the development of
low-complexity
signal processing methods
became
an area
of great interest~\cite{betzel2018approximate}.
In this context,
data compression techniques
are the main tools
for reducing data dimensionality,
a
necessary requisite for efficient data transmission and storage.
There are several methods for data compression~\cite{pennebaker1992jpeg,jolliffe1986principal}
and they are all based on the same principle of reducing
or removing redundancy from the input data~\cite{sayood2017introduction}.
In particular,
transform-based compression methods
generally
map
the input data into smaller output data~\cite{salomon2004data}.

Among several transforms found in the literature,
the
Kar\-hunen-Lo\`eve transform (KLT)~\cite{karhunen1947under,loeve1948functions,britanak2010discrete}
has the distinction
of being
capable of
completely decorrelate
the input signal in the transform domain~\cite{britanak2010discrete,ochoa2019discrete}.
In fact,
the KLT is the optimal linear transform
that minimizes the mean squared error in data compression
for maximum energy concentration
in a few coefficients
of the output signal.
Although mathematically optimal,
the KLT has limited applicability,
because
its derivation depends on
the covariance matrix of the input data,
thus
precluding
or
hindering
the
development of fast algorithms for its computation.
However,
if the input data is a first-order Markov process
with known correlation coefficient~$\rho$,
then
the associate covariance matrix is
deterministically known
and fast algorithms are possible~\cite{ray1970further}.
Nonetheless,
even when a fast algorithm is possible,
it requires multiplications
by a significant amount of irrational numbers,
increasing
the computational cost
of the algorithm.
To the best of our knowledge,
literature is scarce
in methods devoted
to the efficient implementation of the
KLT~\cite{jain1976fast,reed1994fast,fan2019signal}.
Our approach differs from the literature mainly because we consider first-order Markov signals for different values of $\rho$. In this way, depending on the correlation coefficient of the input signal,
we will have a low-complexity transform.

The KLT is mathematically linked to the DCT~\cite{ahmed1974discrete,britanak2010discrete}.
In fact,
the DCT is itself
an asymptotic approximation for the KLT
when
(i)~the input data is first-order Markovian
and
(ii)~the correlation coefficient of the input signal tends the unity~\cite{ahmed1974discrete}.
In contrast to the KLT,
the definition of DCT does not depend on the input signal, which allows the development of fast algorithms computationally efficient.
Thus, DCT became widely adopted
in image and video compression standards such as JPEG~\cite{wallace1992jpeg}, MPEG~\cite{puri2004video}, and HEVC~\cite{pourazad2012hevc}, for example.
However,
even considering state-of-the-art algorithms,
the computational cost of the DCT can still be prohibitive in
scenarios
of
very low processing power or severe restrictions
of energy autonomy~\cite{cintra2014low,bouguezel2008low}.
In this context,
several multiplierless approximations for
the DCT have been proposed~\cite{cintra2014low,bouguezel2008low,haweel2001new,cintra2011dct,potluri2014improved,bayer2012dct,jridi2014generalized,da2017multiplierless,oliveira2019low,canterle2020multiparametric,singhadia2019novel,puchala2021approximate,zidani2019low,chen2019hardware,huang2019deterministic,coelho2021scaling}.
In particular, a widely used approach is to derive approximations based on integer functions~\cite{cintra2014low},
such as the signum and the rounding functions.
Such methodology is employed
to derive
the \emph{signed DCT} (SDCT)~\cite{haweel2001new}
and
the \emph{rounded DCT} (RDCT)~\cite{cintra2011dct}.
In this paper,
we adopt and expand the round-off-based approximation methodology
to the KLT case.
We
aim, therefore,
at the proposition
of a new class of KLT approximations.
Since the transforms are defined deterministically for values of $\rho$ in a predetermined range,
our approach addresses the base exchange problem of the KLT.

This paper is structured as follows. In Section~\ref{S:RKLT}, we present the
mathematical formulation of the KLT for ﬁrst-order Markov data,
the approximation theory, and the design methodology for the proposed approximations.
In Section~\ref{s:results}, the proposed transforms are presented, as well as its assessment measurements.
In Subsection~\ref{ss:fastalgo}, the fast algorithms of the proposed transforms
are displayed. Section~\ref{S:compressao} presents the experiments on image compression and Section~\ref{S:conclusions} concludes the paper.

\section{KLT and Approximate Transforms}\label{S:RKLT}
\subsection{KLT for First-Order Markov Signal}
Let
$\mathbf{x} = [x_0 \quad x_1 \quad \ldots \quad x_{N-1}]^\top$
be an $N$-point random vector.
The Karhunen-Lo\`eve transform
is an $N\times N$ matrix
$\mathbf{K}_N^\rho$
that
maps
$\mathbf{x}$ into the $N$-point uncorrelated vector
$\mathbf{y} = [y_0 \quad y_1 \quad \ldots \quad y_{N-1}]^\top$ given by:
\begin{eqnarray*}
	\mathbf{y} = \mathbf{K}_N^\rho \cdot \mathbf{x}.
\end{eqnarray*}
If $\mathbf{x}$ is a first-order Markov signal, then it was shown in~\cite{jain1976fast} that
the $(i,j)$th entry of the KLT matrix for a given value of the correlation coefficient $\rho \in [0,1]$ is~\cite{britanak2010discrete}:
\begin{equation*}\label{eq:u}
	k_{i,j} = \sqrt{\frac{2}{N+ \lambda_i}} \sin \PR{\omega_i \PC{i - \frac{N-1}{2}}+ \frac{(j+1)\pi}{2} },
\end{equation*}
where
$i,j = 0,1,\ldots,N-1$,
\begin{equation*}\label{eq:lambda}
	\lambda_i = \frac{1-\rho^2}{1+\rho^2 -2\rho \cos \omega_i}
	,
\end{equation*}
and
$\omega_1, \omega_2, \ldots, \omega_{N}$ are the solutions to
\begin{equation*}\label{eq:omega}
	\tan N \omega = \frac{-(1-\rho^2)\sin \omega}{(1+\rho^2)\cos \omega - 2\rho}
	.
\end{equation*}

Since the implementation of the KLT requires float\-ing-point arithmetic, its use has become impractical in real-time applications.
In this context, low-complexity approximations for the KLT are viable solutions to circumvent this problem, since its elements generally require only trivial multiplications and bit-shifting operations.

\subsection{Approximation Theory}

Generally, an approximation is a transform $\mathbf{\widehat{T}}$ that behaves similarly to the exact transform according to some specified figure of merit.
The design of approximate
transforms often requires
the approximations to be orthogonal~\cite{britanak2010discrete}.
Indeed,
if
a matrix is orthogonal,
then its inverse
is equal to its transpose,
and its inverse
is ensured to possess low complexity.
However,
finding orthogonal low-complexity matrices
is not always an easy task.
In~\cite{cintra2014low},
it was shown that if $\mathbf{T}$ is a low-complexity matrix, we can obtain $\widehat{\mathbf{T}}$
through
the polar decomposition~\cite{higham2008functions}.
Thus,
we have that
$\widehat{\mathbf{T}} = \mathbf{S}\cdot \mathbf{T}$,
where
\begin{eqnarray}
	\label{eq:diag}
	\mathbf{S} =
	\begin{cases}
		\sqrt{(\mathbf{T}\cdot\mathbf{T}^\top)^{-1}}, &  \text{if $\mathbf{T}$ is orthogonal,} \label{eq:orto}
		\\
		\sqrt{[\operatorname{diag}(\mathbf{T}\cdot\mathbf{T}^\top)]^{-1}}, &  \text{if $\mathbf{T}$ is non-orthogonal,} \label{eq:no-orto}
	\end{cases}
\end{eqnarray}
and
$\sqrt{\cdot}$
represents the matrix square root operator~\cite{higham2008functions}.
Because
$\mathbf{S}$ is a diagonal matrix, the computational complexity of $\widehat{\mathbf{T}}$ is the same as that of $\mathbf{T}$, except for the multipliers contained in $\mathbf{S}$.
However, the complexity of $\mathbf{S}$ can be absorbed into other sections of a larger procedure, such as the quantization step in the context of image and video compression~\cite{bayer2012dct,bouguezel2011low,lengwehasatit2004scalable,bouguezel2008low,cintra2011dct,bayer2012digital}.
In such cases,
$\mathbf{S}$
does not contribute to
the computational cost \cite{blahut2010fast,britanak2010discrete}.
In the Appendix, we provide a brief derivation showing how the matrix $\mathbf{S}$ can be absorbed into the quantization matrix.

\subsection{Design Methodology}

In a similar fashion as introduced in~\cite{cintra2011dct},
the low-com\-ple\-xity matrix
associated with the rounded KLT (RKLT)
is proposed
according to
the following expression:
\begin{align}
	\mathbf{T}
	\triangleq
	\operatorname{round}(\alpha \cdot \mathbf{K}^{(\rho)}_N)
	,
\end{align}
where $\alpha$ is an expansion factor~\cite{britanak2010discrete},
$\mathbf{K}^{(\rho)}_N$
is the $N$-point KLT matrix for a first-order Markov signal with a given correlation coefficient $0<\rho<1$,
and
the function
$\operatorname{round}(x) =  \lfloor x + 0.5\rfloor$,
with
$\lfloor x \rfloor = \operatorname{max}\{m \in \mathbb{Z} | m \leq x\}$.
When applied to a matrix, the round function operates
element-wise.
To ensure the low-complexity of $\mathbf{T}$,
we restrict its entries to the set $\{0, \pm 1  \}$.
Therefore,
$\alpha$
must satisfy the inequality:
$0 \leq \operatorname{round}(\alpha \cdot \gamma) \leq 1$,
where $\gamma$ is the absolute value of the largest element of the matrix $\mathbf{K}^{(\rho)}_N$.
Thus, we have
$\alpha \in [0,3/2\gamma]$
.

\section{Proposed Approximations}
\label{s:results}

To numerically derive the proposed RKLT transforms,
we adopted the procedure presented in Algorithm~\ref{t:algo}.

\begin{algorithm}
	\caption{Pseudocode for deriving low-complexity matrices.}
	\label{t:algo}
	\begin{algorithmic}[]
		\renewcommand{\algorithmicrequire}{\textbf{Input:}}
		\renewcommand{\algorithmicensure}{\textbf{Output:}}
		\REQUIRE{$N$, $\alpha$, step}
		\ENSURE{Set $\mathcal{C}$ of low-complexity matrices}
		\STATE $\mathcal{C} = \{\}$,
		$\mathbf{T} = \mathbf{0}_{N\times N}$;
		\FOR{$\rho = \text{step}: \text{step}:(1 - \text{step})$}
		\STATE $\mathbf{T}^\prime :=
		\operatorname{round}(\alpha \cdot \mathbf{K}^{(\rho)}_N)$;
		\IF{$\mathbf{T}^\prime \neq \mathbf{T}$}
		\STATE $\mathcal{C} := \mathcal{C} \cup \{\mathbf{T}^\prime\}$;
		\STATE $\mathbf{T} := \mathbf{T}^\prime$;
		\ENDIF
		\ENDFOR

		\noindent
		\RETURN $\mathcal{C}$
	\end{algorithmic}
\end{algorithm} \color{black}
The extreme values $\rho = 0$ and $\rho = 1$ were not considered,
because
they result, respectively,
in a degenerate covariance matrix
and
in the exact DCT matrix,
whose approximation theory is covered in~\cite{britanak2010discrete}.
Such methodology is capable of finding approximation
for any blocklength $N$.
In this paper, we focus on the case $N = 8$
due to its wide significance
in the image and video coding.
In this case, the range of $\alpha$ is approximately $[0,3.07]$.
Thus, we adopted $\alpha = 2$, agreeing with the methodology~\cite{cintra2011integer}
employed to derive the RDCT~\cite{cintra2011dct}.
Exact KLT matrices were obtained for values of $\rho \in [0.1,0.9]$ with steps of $10^{-1}$.
Table~\ref{t:RKLTS}
presents the obtained transforms and their respective diagonals $\mathbf{S}$,
as well as the intervals of $\rho$.
Matrix $\mathbf{T}_4$
coincides
with the transformation shown in~\cite{cintra2011dct}.

\begin{table*}[h!] \caption{RKLT approximations}
	\label{t:RKLTS}
	\centering
	\begin{tabular}{lccccc}
		\toprule
		Transform &  $\rho$ & Matrix
		& $\mathbf{S}$ \\
		\midrule
		\addlinespace[1.5ex]
		$\mathbf{T}_1$ %
		& $(0,0.4)$ &
		$\left[ \begin{rsmallmatrix}
			0 & 1 & 1 & 1 & 1 & 1 & 1 & 0 \\
			1 & 1 & 1 & 0 & 0 & -1 & -1 & -1 \\
			1 & 1 & 0 & -1 & -1 & 0 & 1 & 1 \\
			1 & 0 & -1 & -1 & 1 & 1 & 0 & -1 \\
			1 & 0 & -1 & 1 & 1 & -1 & 0 & 1 \\
			1 & -1 & 0 & 1 & -1 & 0 & 1 & -1 \\
			1 & -1 & 1 & 0 & 0 & 1 & -1 & 1 \\
			0 & -1 & 1 & -1 & 1 & -1 & 1 & 0 \\
		\end{rsmallmatrix} \right]$
		& $\operatorname{diag} \PC{\frac{1}{\sqrt{6}},\frac{1}{\sqrt{6}},\frac{1}{\sqrt{6}},\frac{1}{\sqrt{6}},\frac{1}{\sqrt{6}},\frac{1}{\sqrt{6}},\frac{1}{\sqrt{6}},\frac{1}{\sqrt{6}}}$
		\\
		\addlinespace[1.5ex]
		$\mathbf{T}_2$ %
		& $[0.4,0.7)$ &
		$\left[
		\begin{rsmallmatrix}
			0 & 1 & 1 & 1 & 1 & 1 & 1 & 0 \\
			1 & 1 & 1 & 0 & 0 & -1 & -1 & -1 \\
			1 & 1 & 0 & -1 & -1 & 0 & 1 & 1 \\
			1 & 0 & -1 & -1 & 1 & 1 & 0 & -1 \\
			1 & -1 & -1 & 1 & 1 & -1 & -1 & 1 \\
			1 & -1 & 0 & 1 & -1 & 0 & 1 & -1 \\
			0 & -1 & 1 & 0 & 0 & 1 & -1 & 0 \\
			0 & -1 & 1 & -1 & 1 & -1 & 1 & 0 \\
		\end{rsmallmatrix}
		\right]$
		& $\operatorname{diag} \PC{\frac{1}{\sqrt{6}},\frac{1}{\sqrt{6}},\frac{1}{\sqrt{6}},\frac{1}{\sqrt{6}},\frac{1}{2\sqrt{2}},\frac{1}{\sqrt{6}},\frac{1}{2},\frac{1}{\sqrt{6}}}$ \\
		\addlinespace[1.5ex]
		$\mathbf{T}_3$ %
		&  $[0.7,0.8)$ &
		$\left[
		\begin{rsmallmatrix}
			1 & 1 & 1 & 1 & 1 & 1 & 1 & 1 \\
			1 & 1 & 1 & 0 & 0 & -1 & -1 & -1 \\
			1 & 1 & 0 & -1 & -1 & 0 & 1 & 1 \\
			1 & 0 & -1 & -1 & 1 & 1 & 0 & -1 \\
			1 & -1 & -1 & 1 & 1 & -1 & -1 & 1 \\
			1 & -1 & 0 & 1 & -1 & 0 & 1 & -1 \\
			0 & -1 & 1 & 0 & 0 & 1 & -1 & 0 \\
			0 & -1 & 1 & -1 & 1 & -1 & 1 & 0 \\
		\end{rsmallmatrix}
		\right]$
		& $\operatorname{diag} \PC{\frac{1}{2\sqrt{2}},\frac{1}{\sqrt{6}},\frac{1}{\sqrt{6}},\frac{1}{\sqrt{6}},\frac{1}{2\sqrt{2}},\frac{1}{\sqrt{6}},\frac{1}{2},\frac{1}{\sqrt{6}}}$ \\
		\addlinespace[1.5ex]
		$\mathbf{T}_4$~\cite{cintra2011dct} %
		& $[0.8,1)$ &
		$\left[
		\begin{rsmallmatrix}
			1 & 1 & 1 & 1 & 1 & 1 & 1 & 1 \\
			1 & 1 & 1 & 0 & 0 & -1 & -1 & -1 \\
			1 & 0 & 0 & -1 & -1 & 0 & 0 & 1 \\
			1 & 0 & -1 & -1 & 1 & 1 & 0 & -1 \\
			1 & -1 & -1 & 1 & 1 & -1 & -1 & 1 \\
			1 & -1 & 0 & 1 & -1 & 0 & 1 & -1 \\
			0 & -1 & 1 & 0 & 0 & 1 & -1 & 0 \\
			0 & -1 & 1 & -1 & 1 & -1 & 1 & 0 \\
		\end{rsmallmatrix}
		\right]$
		& $\operatorname{diag} \PC{\frac{1}{2\sqrt{2}},\frac{1}{\sqrt{6}},\frac{1}{2},\frac{1}{\sqrt{6}},\frac{1}{2\sqrt{2}},\frac{1}{\sqrt{6}},\frac{1}{2},\frac{1}{\sqrt{6}}}$ \\
		\addlinespace[1.5ex]
		\bottomrule
	\end{tabular}
\end{table*}

\subsection{Assessment Metrics}

To evaluate the performance of the proposed transforms,
we
considered
two types of
figures of merit:
(i)~coding measures,
such as the unified coding gain~\cite{katto1992short}
and transform efficiency~\cite{nikara2001unified},
which measure decorrelation and energy compaction;
and
(ii)~proximity measures,
such as the mean square error~\cite{britanak2010discrete}
and total error energy~\cite{cintra2011dct},
which measure similarities between
approximate and exact matrices
in a Euclidean distance sense.
We detailed each of these measures below.
\subsubsection{Unified Coding Gain}\label{sss:cg}
The unified coding gain quantifies the energy compaction capability of the transform $\widehat{\mathbf{T}}$ and is given by~\cite{katto1992short}:
\begin{eqnarray*}
{\textrm{Cg}}(\widehat{\mathbf{T}}) =
10 \cdot
\log_{10}
\Biggl\{
\prod_{k=1}^N
\frac{1}{\sqrt[N]{A_k \cdot B_k}}
\Biggr\}
,
\end{eqnarray*}
where
$
A_k
=
\operatorname{su}
\left\{
(\mathbf{h}_k^\top \cdot \mathbf{h}_k)\odot \mathbf{R_x}
\right\}
$,
$\mathbf{h}_k$ is the $k$th
row
vector from $\widehat{\mathbf{T}}$,
the function $\operatorname{su}(\cdot)$ returns the sum of the elements of its matrix argument, $\odot$ is the Hadamard matrix product operator~\cite{seber2008matrix},
$\mathbf{R_x}$ is the autocorrelation matrix of the considered first-order Markov signal,
$B_k = \| \mathbf{g}_k \|^2$,
$\mathbf{g}_k$ is the $k$th
row vector from~$\widehat{\mathbf{T}}^{-1}$, and $\| \cdot \|$ is the Frobenius norm~\cite{seber2008matrix}.

\subsubsection{Transform Efficiency}\label{sss:eta}
The transform efficiency is another coding related figure of merit, and is given by~\cite{nikara2001unified}:
\begin{eqnarray*}
	\eta({\widehat{\mathbf{T}}}) = 100\cdot \frac{ \sum_{i = 1}^{N} \vert r_{i,i}\vert}{ \sum_{i = 1}^{N}  \sum_{j = 1}^{N} \vert r_{i,j} \vert},
\end{eqnarray*}
where $r_{i,j}$ is the $(i,j)$th element from
$\widehat{\mathbf{T}}\cdot
\mathbf{R_x}
\cdot \widehat{\mathbf{T}}^\top$.

\subsubsection{Mean Square Error}\label{sss:mse}
The mean square error (MSE) relative to the exact KLT is given by~\cite{britanak2010discrete}:
\begin{eqnarray*}
	\textrm{MSE}({\widehat{\mathbf{T}}}) = \frac{1}{N}\cdot
	\operatorname{tr}
	\left\{
	(\mathbf{K}^{(\rho)}_N - {\widehat{\mathbf{T}}})\cdot \mathbf{R_x} \cdot (\mathbf{K}^{(\rho)}_N - {\widehat{\mathbf{T}}})^\top
	\right\}
	,
\end{eqnarray*}
where
$\operatorname{tr}(\cdot)$ is the trace function~\cite{harville1997trace}.

\subsubsection{Total Error Energy}\label{sss:error}
Another error measure is the total error energy, which is given by~\cite{cintra2011dct}:
\begin{eqnarray*}
	\epsilon(\widehat{\mathbf{T}}) = \pi \cdot \|\mathbf{K}^{(\rho)}_N - {\widehat{\mathbf{T}}}\|^2.
\end{eqnarray*}

Strictly there is no competing method in the literature that could allow us to make a fair comparison. Thus, we compared
the proposed approximation only
with
the exact KLT for $\rho = 0.3$, $0.4$, $0.7$, and $0.8$.
We considered the exact same values of $\rho$ for computing the unified coding gain and the
transform efficiency, which depends on the autocorrelation matrix $\mathbf{R_x}$ of the
considered first-order Markov input signal.
Thus, we assessed $\widehat{\mathbf{T}}_1$ compared to the $\mathbf{K}^{(0.3)}$ for the value of $\rho = 0.3$, $\widehat{\mathbf{T}}_2$ compared to the $\mathbf{K}^{(0.4)}$ for the value of $\rho = 0.4$, $\widehat{\mathbf{T}}_3$ compared to the $\mathbf{K}^{(0.7)}$ for the value of $\rho = 0.7$, and $\widehat{\mathbf{T}}_4$ compared to the $\mathbf{K}^{(0.8)}$ for the value of $\rho = 0.8$.

To the best of our knowledge,
the literature
lacks
efficient KLT-based methods
for
lowly correlated data.
We aim at contributing to filling this gap. \color{black}

Table~\ref{t:measuresRKLT8}
presents
the coding and similarity
measurements.
As reference values,
the total error energy
and
mean square error
for the RDCT~\cite{cintra2011dct} compared to the DCT (for $\rho = 0.95$)
are, respectively, $1.7945$ and $0.0098$.
Note that for the proximity measures (mean square error and total error energy),
the proposed transforms perform well, even better than the performance of the RDCT compared to the DCT considering the total error energy. In this case, the smaller the measurement is, the more similar the approximate transform is to the exact one.
Considering the coding measures (unified coding gain and transform efficiency),
the proposed transforms assert their good performances
when compared with the measurements from the exact KLT for each value of $\rho$. The exact KLT coding measurements have been used as a benchmark to evaluate the performance of other transforms since it is the unitary optimal transform in terms of energy compaction and decorrelation~\cite{ochoa2019discrete,britanak2010discrete}.
It is notable that the performance of the proposed approximations are similar to the exact KLT and have a greatly reduced computational cost, as detailed in the following.
\color{black}

\begin{table*}[t] \caption{Coding and similarity measures}
	\label{t:measuresRKLT8}
	\centering
	\begin{tabular}{cccccc}
		\toprule
		& $\rho$  & ${\textrm{Cg}}(\widehat{\mathbf{T}})$ & $\eta(\widehat{\mathbf{T}})$ & $\epsilon(\mathbf{K}^{(\rho)}, \widehat{\mathbf{T}})$ & ${\textrm{MSE}}(\mathbf{K}^{(\rho)}, \widehat{\mathbf{T}})$\\
		\midrule
		$\mathbf{K}^{(0.3)}$ & $0.3$&$0.3584$&$ 100 $&$0$&$ 0$\\
		$\widehat{\mathbf{T}}_1$ & $(0,0.4)$&$0.2829$&$80.7088$&$1.6751$&$0.0659$\\
		\addlinespace[1.5ex]
		\midrule
		\addlinespace[1.5ex]

		$\mathbf{K}^{(0.4)}$ & $0.4$&$0.6626$&$ 100 $&$0$&$ 0$\\
		$\widehat{\mathbf{T}}_2$ & $[0.4,0.7)$&$0.5616$&$70.2996$&$1.7011$&$0.0660$\\
		\addlinespace[1.5ex]
		\midrule
		\addlinespace[1.5ex]
		$\mathbf{K}^{(0.7)}$ & $0.7$&$2.5588$&$ 100 $&$0$&$ 0$\\

		$\widehat{\mathbf{T}}_3$ & $[0.7,0.8)$&$2.1398$&$65.8777$&$1.4716$&$0.0523$\\
		\addlinespace[1.5ex]
		\midrule
		\addlinespace[1.5ex]
		$\mathbf{K}^{(0.8)}$ & $0.8$&$3.8824$&$ 100 $&$0$&$ 0$\\

		$\widehat{\mathbf{T}}_4$~\cite{cintra2011dct} & $[0.8,1)$&$3.4058$&$74.4747$&$1.7715$&$0.0362$\\
		\bottomrule
	\end{tabular}
\end{table*}

\subsection{Fast Algorithm and Computational Complexity}\label{ss:fastalgo}

\begin{figure}[]
	\centering
	\includegraphics[width=8cm]{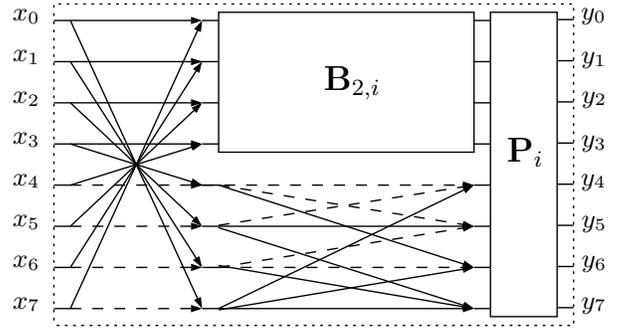}
	\caption{SFG of the proposed transforms. Block $\mathbf{B}_{2,i}$ is different for each transform and it is presented in Fig.~\ref{f:SFGs}.} \label{f:SFG}
\end{figure}

Fast algorithms for the approximate transforms
can be derived
based on
the sparse factorization
of the transform matrices
and
butterfly matrix structures~\cite{blahut2010fast}.
The factorizations of the proposed transform are given by:
\begin{eqnarray*}
	\mathbf{T}_i = \mathbf{P}_i \cdot \mathbf{A}_{2,i} \cdot \mathbf{A}_1
	\quad
	\text{for} \quad i = 1,2,3,4,
\end{eqnarray*}
where
\begin{eqnarray*}
	\label{eq:A1}
	\mathbf{A}_1 =
	\begin{bmatrix}
		\mathbf{I}_4 & \bar{\mathbf{I}}_4 \\
		\bar{\mathbf{I}}_4 & -\mathbf{I}_4
	\end{bmatrix}, \quad
	\mathbf{A}_{2,i} =
	\begin{bmatrix}
		\mathbf{B}_{2,i} & \\
		& \mathbf{B}_2
	\end{bmatrix},
\end{eqnarray*}
\begin{eqnarray*}
	\mathbf{B}_2 =
	\PR{
		\begin{matrix}
			-1 & -1 & 0 & 1 \\
			-1 & 1 & -1 & 0 \\
			1 & 0 & -1 & 1 \\
			0 & 1 & 1 & 1
	\end{matrix}},
\end{eqnarray*}
and $\mathbf{I}_4 $ and $\bar{\mathbf{I}}_4$ are, respectively, the identity and counter-identity matrices of order $4$. Matrices $\mathbf{B}_{2,i}$ are given by:
\begin{eqnarray*}
	\mathbf{B}_{2,1} =
	\PR{
		\begin{matrix}
			1 & 1 & 0 & -1 \\
			1 & -1 & 1 & 0 \\
			1 & 0 & -1 & 1 \\
			0 & 1 & 1 & 1
	\end{matrix}}, \quad
	\mathbf{B}_{2,2} =
	\PR{
		\begin{matrix}
			1 & 1 & 0 & -1 \\
			1 & -1 & -1 & 1 \\
			0 & -1 & 1 & 0 \\
			0 & 1 & 1 & 1
	\end{matrix}},
\end{eqnarray*}
\begin{eqnarray*}
	\mathbf{B}_{2,3} =
	\PR{
		\begin{matrix}
			1 & 1 & 1 & 0 \\
			1 & -1 & -1 & 0 \\
			0 & -1 & 1 & 0 \\
			0 & 1 & 0 & 1 \\
	\end{matrix}} \cdot
	\PR{
		\begin{matrix}
			1 & 0 & 0 & 1 \\
			0 & 1 & 0 & 0 \\
			0 & 0 & 1 & 0 \\
			1 & 0 & 0 & -1 \\
	\end{matrix}},
\end{eqnarray*}
\begin{eqnarray*}
	\mathbf{B}_{2,4} =
	\PR{
		\begin{matrix}
			1 & 1 & 0 & 0 \\
			1 & -1 & 0 & 0 \\
			0 & 0 & -1 & 0 \\
			0 & 0 & 0 & 1 \\
	\end{matrix}} \cdot
	\PR{
		\begin{matrix}
			1 & 0 & 0 & 1 \\
			0 & 1 & 1 & 0 \\
			0 & 1 & -1 & 0 \\
			1 & 0 & 0 & -1 \\
	\end{matrix}}.
\end{eqnarray*}
The permutation matrices $\mathbf{P}_i$ are:
\begin{align*}
	\mathbf{P}_1 =
		\begin{bmatrix}
			0 & 0 & 0 & 1 & 0 & 0 & 0 & 0 \\
			0 & 0 & 0 & 0 & 0 & 0 & 0 & 1 \\
			1 & 0 & 0 & 0 & 0 & 0 & 0 & 0 \\
			0 & 0 & 0 & 0 & 1 & 0 & 0 & 0 \\
			0 & 0 & 1 & 0 & 0 & 0 & 0 & 0 \\
			0 & 0 & 0 & 0 & 0 & 0 & 1 & 0 \\
			0 & 1 & 0 & 0 & 0 & 0 & 0 & 0 \\
			0 & 0 & 0 & 0 & 0 & 1 & 0 & 0 \\
	\end{bmatrix},
\end{align*}
\begin{align*}
	\mathbf{P}_2 =
		\begin{bmatrix}
			0 & 0 & 0 & 1 & 0 & 0 & 0 & 0 \\
			0 & 0 & 0 & 0 & 0 & 0 & 0 & 1 \\
			1 & 0 & 0 & 0 & 0 & 0 & 0 & 0 \\
			0 & 0 & 0 & 0 & 1 & 0 & 0 & 0 \\
			0 & 1 & 0 & 0 & 0 & 0 & 0 & 0 \\
			0 & 0 & 0 & 0 & 0 & 0 & 1 & 0 \\
			0 & 0 & 1 & 0 & 0 & 0 & 0 & 0 \\
			0 & 0 & 0 & 0 & 0 & 1 & 0 & 0 \\
	\end{bmatrix},
\end{align*}
and
\begin{align*}
	\mathbf{P}_3 = \mathbf{P}_4 =
		\begin{bmatrix}
			1 & 0 & 0 & 0 & 0 & 0 & 0 & 0 \\
			0 & 0 & 0 & 0 & 0 & 0 & 0 & 1 \\
			0 & 0 & 0 & 1 & 0 & 0 & 0 & 0 \\
			0 & 0 & 0 & 0 & 1 & 0 & 0 & 0 \\
			0 & 1 & 0 & 0 & 0 & 0 & 0 & 0 \\
			0 & 0 & 0 & 0 & 0 & 0 & 1 & 0 \\
			0 & 0 & 1 & 0 & 0 & 0 & 0 & 0 \\
			0 & 0 & 0 & 0 & 0 & 1 & 0 & 0 \\
	\end{bmatrix}.
\end{align*}
Fig.~\ref{f:SFG} and \ref{f:SFGs} show the signal flow graphs (SFG)
of the fast algorithms. The dashed arrows represent multiplication by $-1$.

\begin{figure*}[]
	\centering
	\subfigure[\scriptsize $\mathbf{B}_{2,1}$]{\label{f:SFG-K1}\includegraphics[width=6cm]{SFG-K1.eps}}
	\subfigure[\scriptsize $\mathbf{B}_{2,2}$]{\label{f:SFG-K2}\includegraphics[width=6cm]{SFG-K2.eps}}
	\\
	\subfigure[\scriptsize $\mathbf{B}_{2,3}$]{\label{f:SFG-K3}\includegraphics[width=6cm]{SFG-K3.eps}}
	\subfigure[\scriptsize $\mathbf{B}_{2,4}$]{\label{f:SFG-K4}\includegraphics[width=6cm]{SFG-K4.eps}}
	\caption{Block $\mathbf{B}_{2,i}$ of each proposed transform.}\label{f:SFGs}
\end{figure*}

The direct implementation of the exact KLT requires $56$ additions and $64$ multiplications.
The proposed transforms are designed to be multiplierless but still require $56$ additions in its direct implementation.
Considering the fast algorithms proposed for the transforms, we have a reduction in the addition operations relative to the direct implementation.
Matrices $\mathbf{T}_1$, $\mathbf{T}_2$, and $\mathbf{T}_3$ require only $24$ additions, causing a reduction of $57.17 \%$ in the number of addition o\-pe\-rations, and matrix $\mathbf{T}_4$ requires $22$ additions, with a reduction of $60.71 \%$.

\section{Image Compression}
\label{S:compressao}

In this section,
the performance of the proposed transforms
is assessed in
the context of image compression
~\cite{gonzalez2002digital}
as suggested in~\cite{cintra2011dct,bouguezel2008low,cintra2014low,haweel2001new,jridi2015generalized,da2017multiplierless,oliveira2019low}.

\subsection{JPEG-like Compression}
The following compression scheme~\cite{salomon2004data} was applied to
standardized images obtained from
the public image bank available in~\cite{uscsipi}. \color{black}
Input images were divided into disjoint sub-blocks of size $8 \times 8$.
The 2D
direct and inverse transformations
induced by
$\mathbf{K}_N^\rho$
are computed, respectively,
by~\cite{suzuki2010integer}:
\begin{eqnarray*}
	\mathbf{B} =\mathbf{K}_N^\rho \cdot \mathbf{A} \cdot (\mathbf{K}_N^\rho)^{-1}, \\
	\mathbf{A} = (\mathbf{K}_N^\rho)^{-1} \cdot \mathbf{B} \cdot \mathbf{K}_N^\rho
	,
\end{eqnarray*}
where $\mathbf{A}$ and $\mathbf{B}$
are square matrices of size $N$.
Each sub-block was submitted to the 2D transform computation
and the resulting transform-domain coefficients
were re-ordered using the standard zig-zag sequence~\cite{salomon2004data}.
Only the initial $r$ coefficients in each sub-block were retained and the remaining  coefficients were zeroed.
The 2D inverse transform was applied and
the reconstructed sub-blocks were adequately rearranged.
Original and
compressed images were then evaluated considering
traditional quality assessment measures.
The considered figures of merit for image quality evaluation
were:
(i)~the mean structural similarity index (MSSIM)~\cite{wang2004image};
(ii)~the mean square error (MSE)~\cite{britanak2010discrete};
(iii)~and the peak signal-to-noise ratio (PSNR)~\cite{Huynh-Thu2008Scope}.
Even though the PSNR and MSE
are very popular figures of merit, it was shown in~\cite{wang2009mean} that it might offer
limited results as image quality assessment tools
for it poorly correlates with human perception. \color{black}
On the other hand,
the MSSIM is capable of closely capturing
the image quality as understood by the human visual system model~\cite{wang2004image}.
The image compression experiments were divided into two analyses: (i)~a qualitative one, based on
compressed \textit{Lena}, \textit{Baboon}, and \textit{Moon} images with approximately $77\%$ of compression rate; (ii)~and a quantitative analysis, based on the average measures of $45$ standardized $8$-bit compressed images~\cite{uscsipi} for a wide range of retained coefficients ($r$).
The results are presented below.
\color{black}

\subsection{Results}
\begin{figure*}[h]
	\centering
	\subfigure[\textit{Lena}]{\label{f:lena}
		\includegraphics[width=3.5cm]{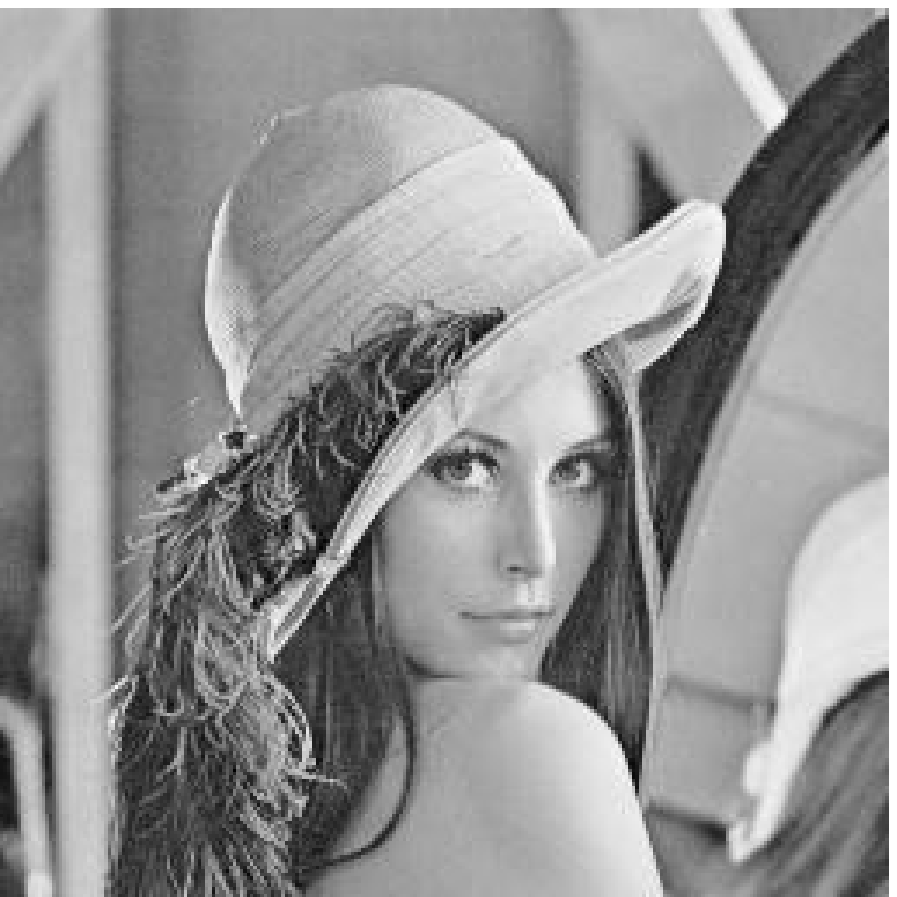}}
	\subfigure[\textit{Baboon}]	{\label{f:baboon}
		\includegraphics[width=3.5cm]{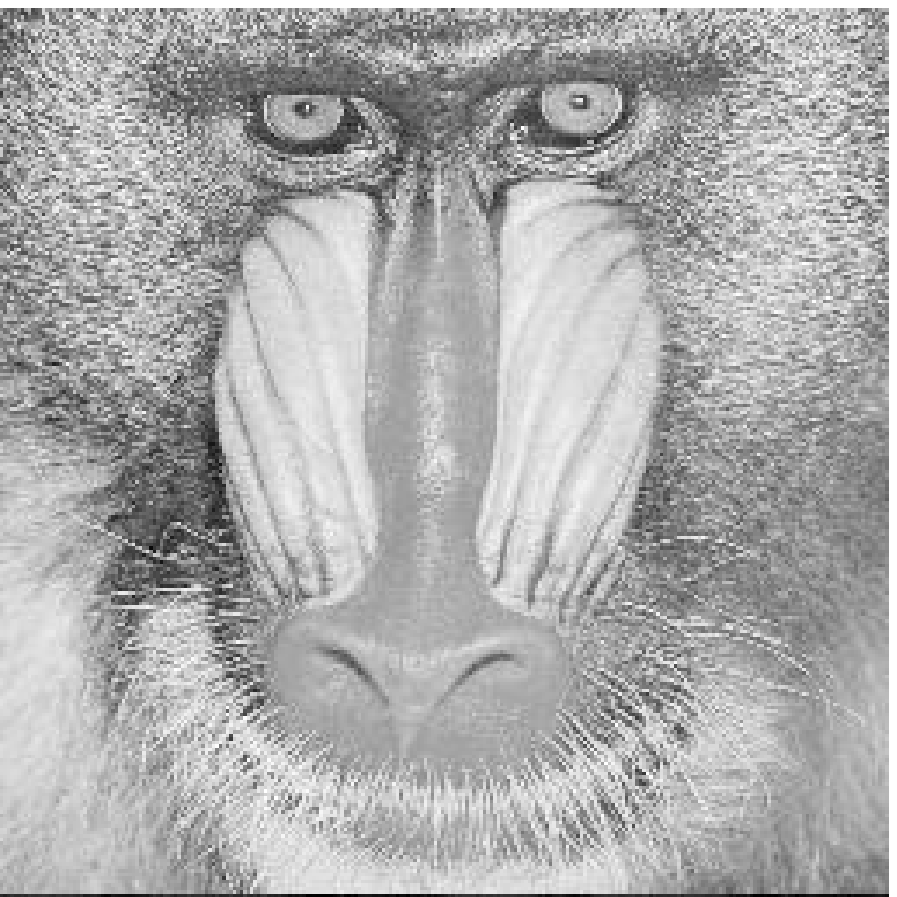}}
	\subfigure[\textit{Moon}] {\label{f:grass}
		\includegraphics[width=3.5cm]{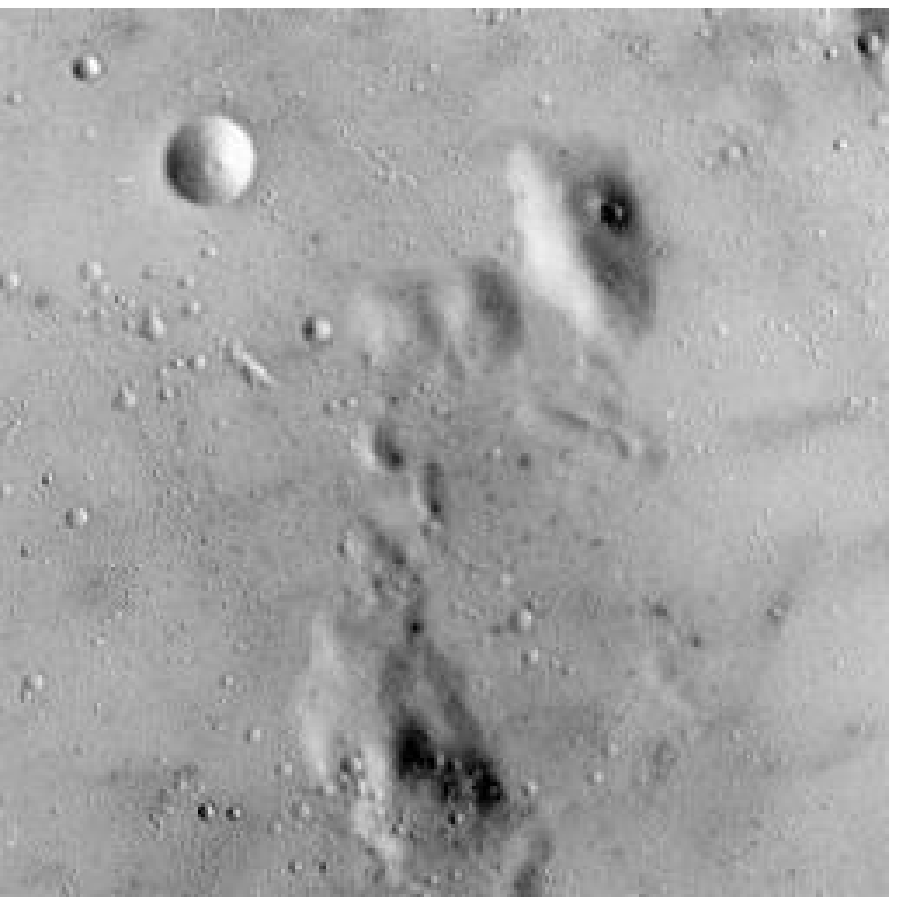}}
	\caption{Original images.}\label{f:Original}
\end{figure*}

For the qualitative analysis, we considered three known public images available on~\cite{uscsipi}.
The original grayscale images are presented in Fig.~\ref{f:Original}.

\begin{figure*}[h!]
	\centering
	\subfigure[$\widehat{\mathbf{T}}_1$]{\label{f:lenaK1}
		\includegraphics[width=4cm]
		{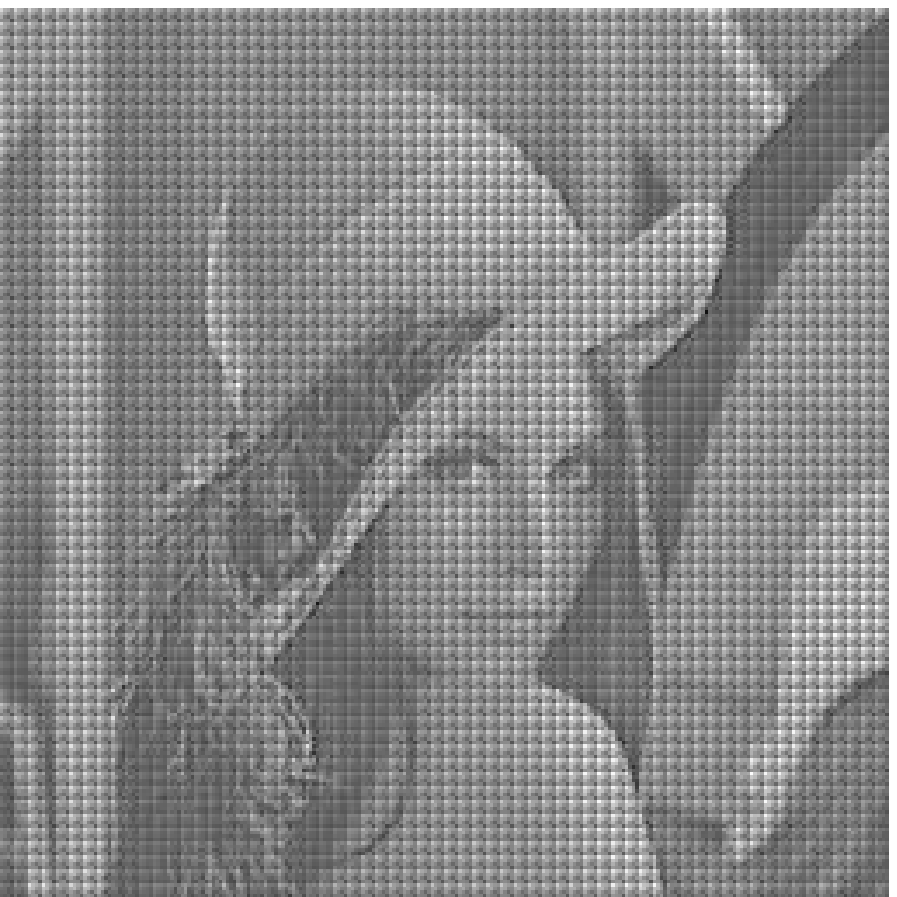}}
	\subfigure[$\widehat{\mathbf{T}}_2$]
	{\label{f:lenaK2}
			\includegraphics[width=4cm]
		{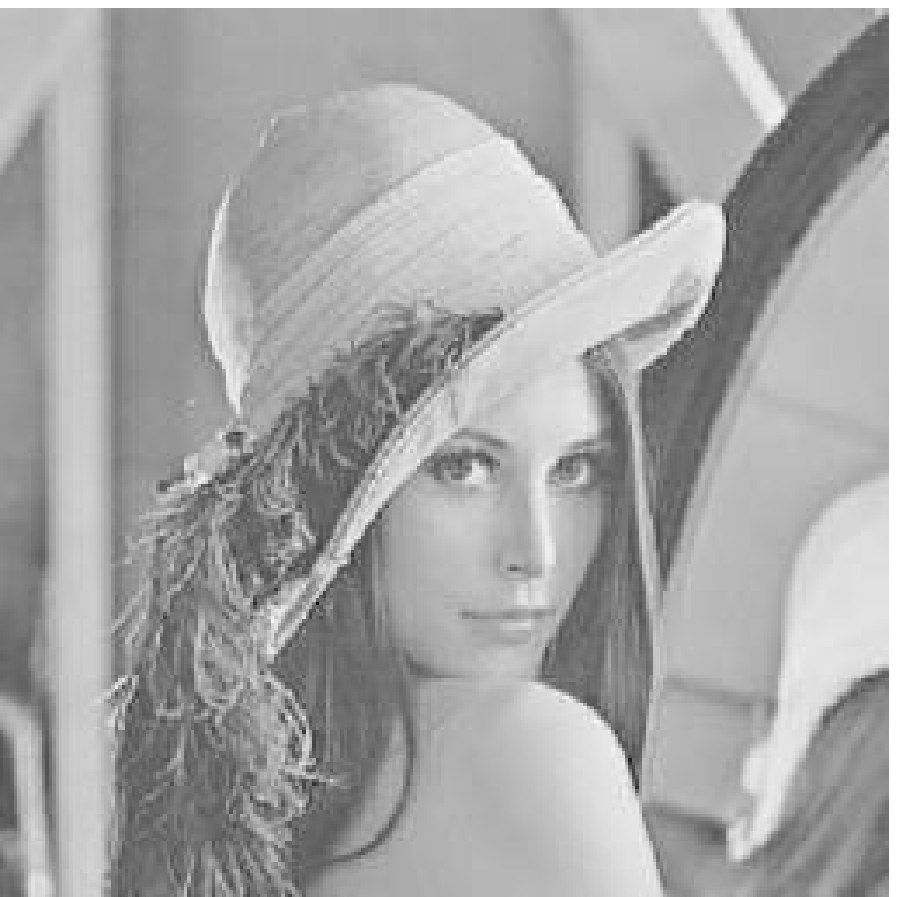}}
	\subfigure[$\widehat{\mathbf{T}}_3$]
	{\label{f:lenaK3}
		\includegraphics[width=4cm]
		{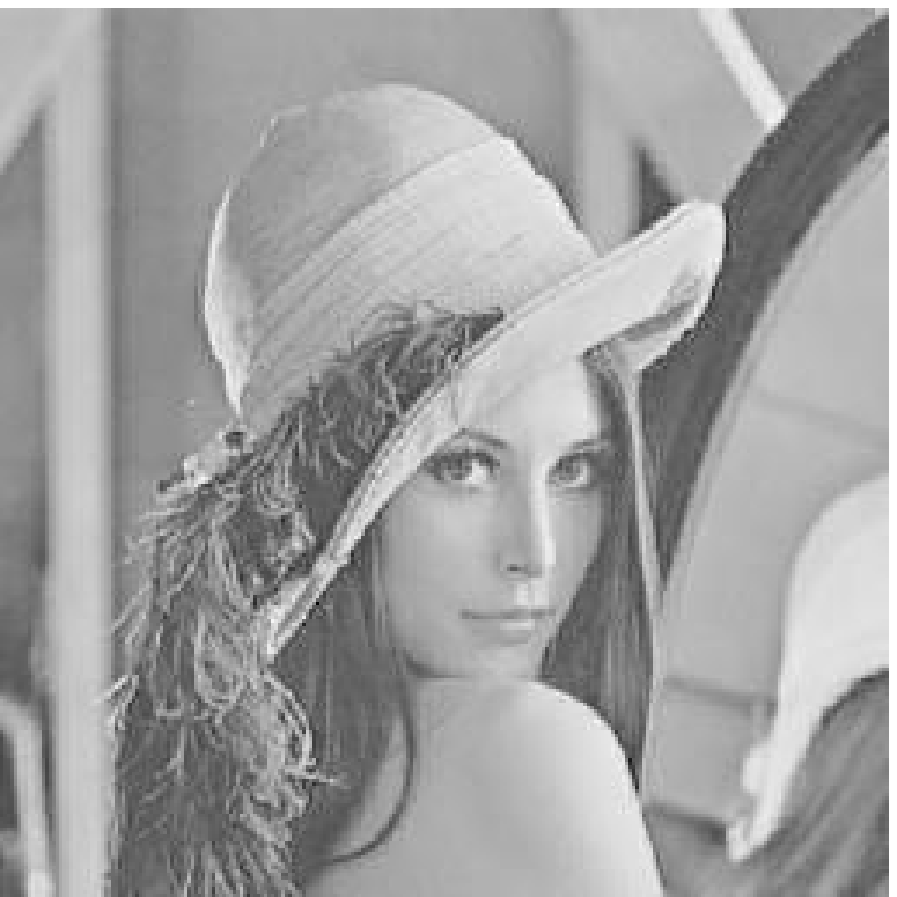}}
	\subfigure[$\widehat{\mathbf{T}}_4$~\cite{cintra2011dct}]{\label{f:lenaK4}
		\includegraphics[width=4cm]
		{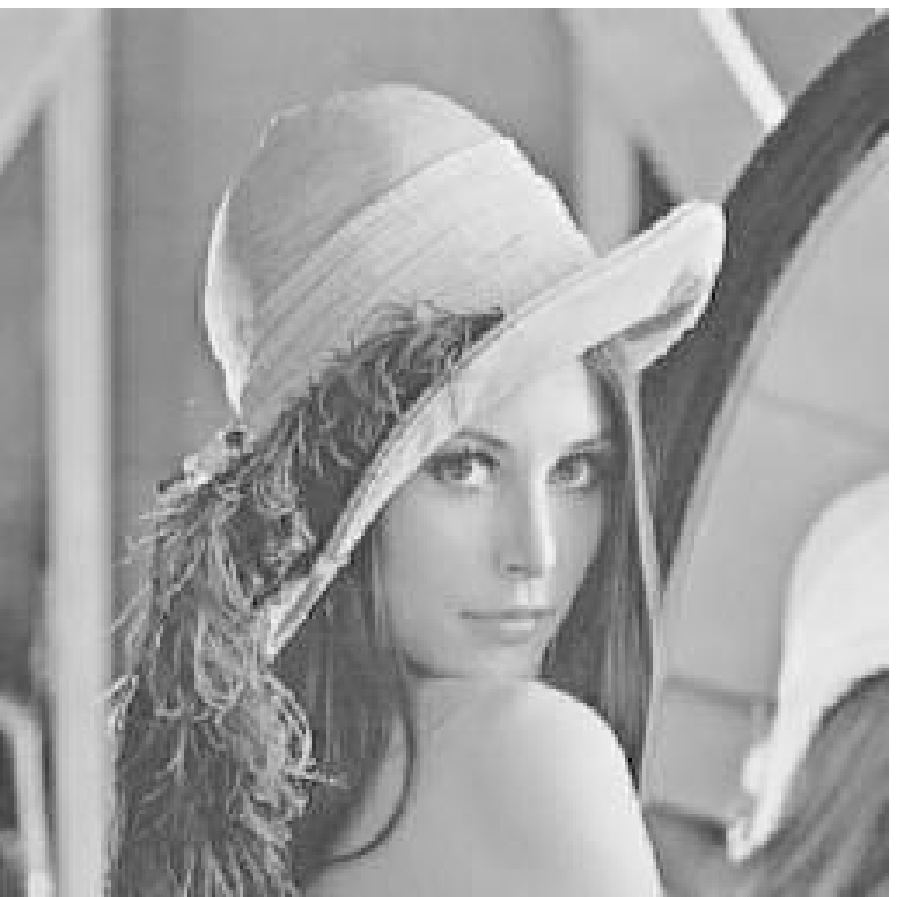}}  \\
	\subfigure[$\mathbf{K}^{(0.3)}$]{\label{f:lenaK03}
		\includegraphics[width=4cm]
		{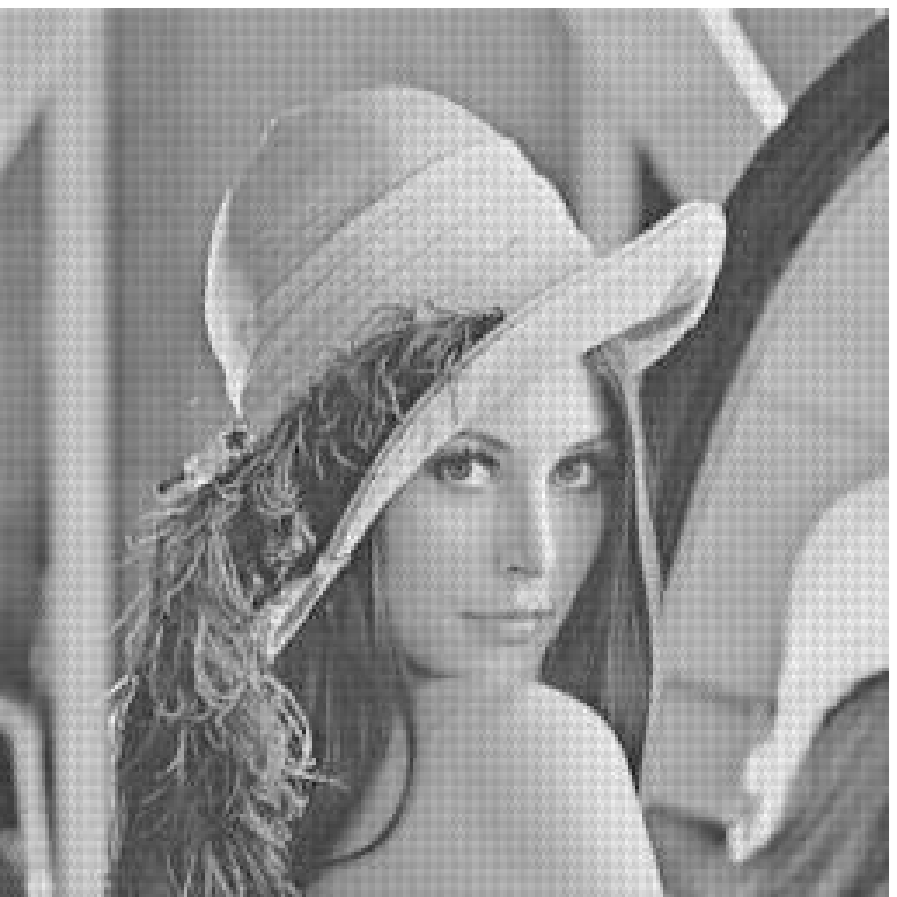}}
	\subfigure[$\mathbf{K}^{(0.4)}$]{\label{f:lenaK04}
		\includegraphics[width=4cm]
		{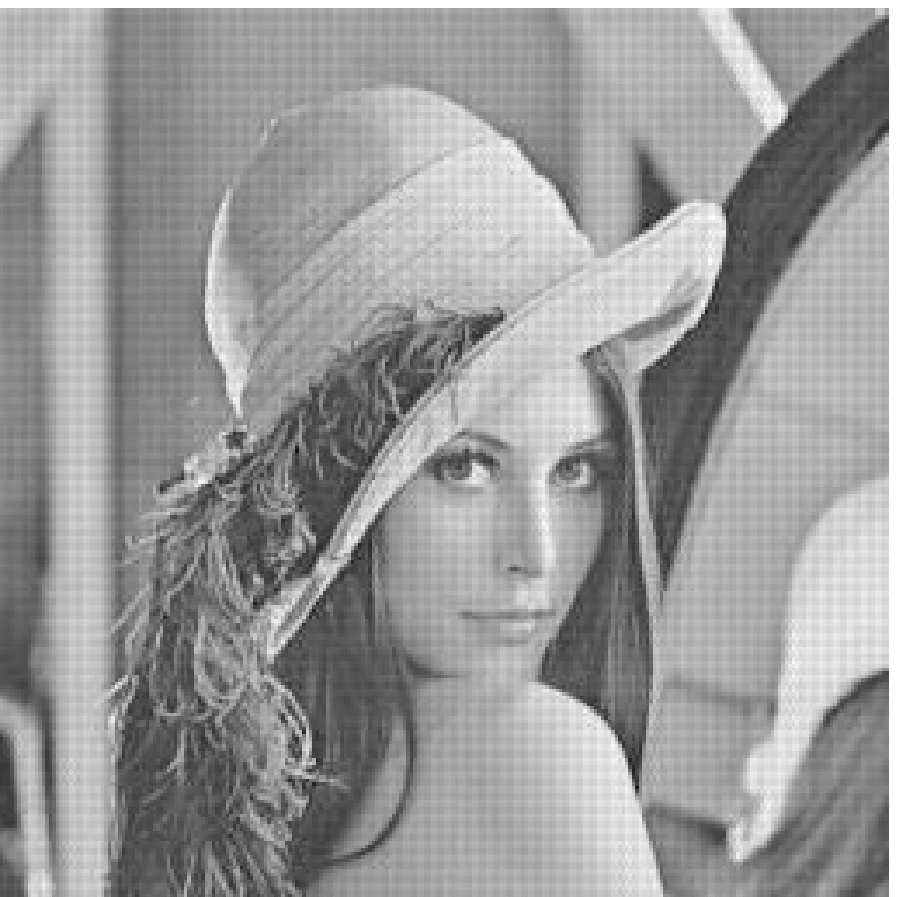}}
	\subfigure[$\mathbf{K}^{(0.7)}$]{\label{f:lenaK07}
		\includegraphics[width=4cm]
		{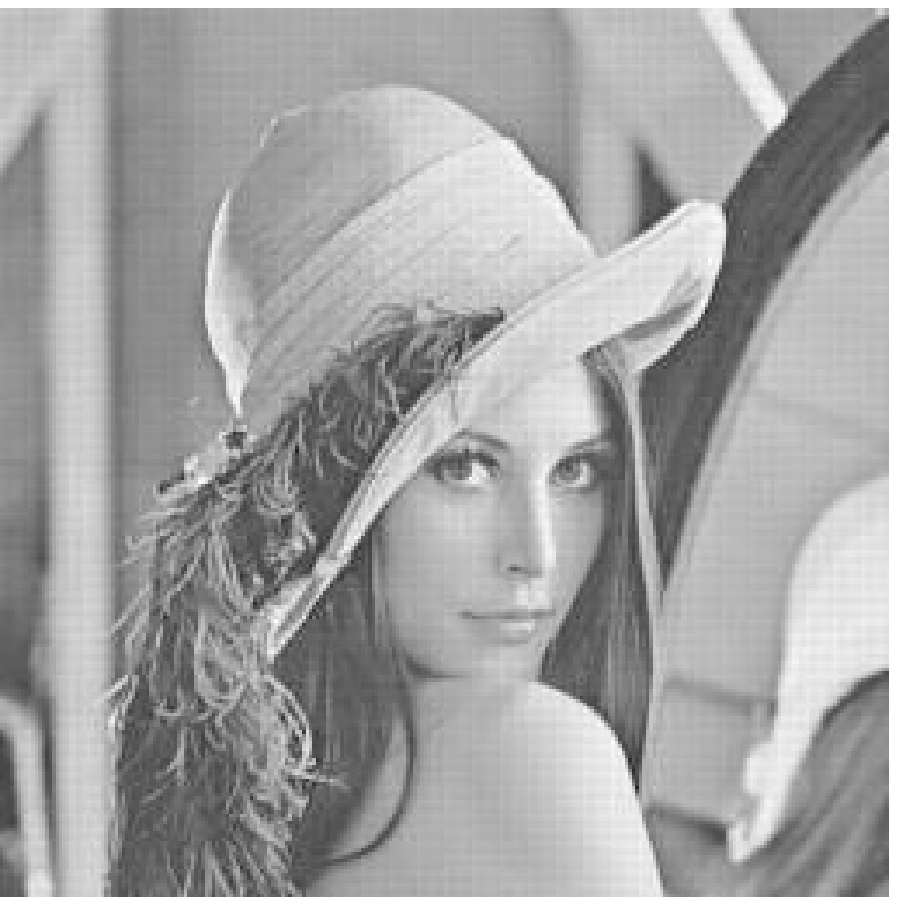}}
		\subfigure[$\mathbf{K}^{(0.8)}$]
	{\label{f:lenaK08}
		\includegraphics[width=4cm]
		{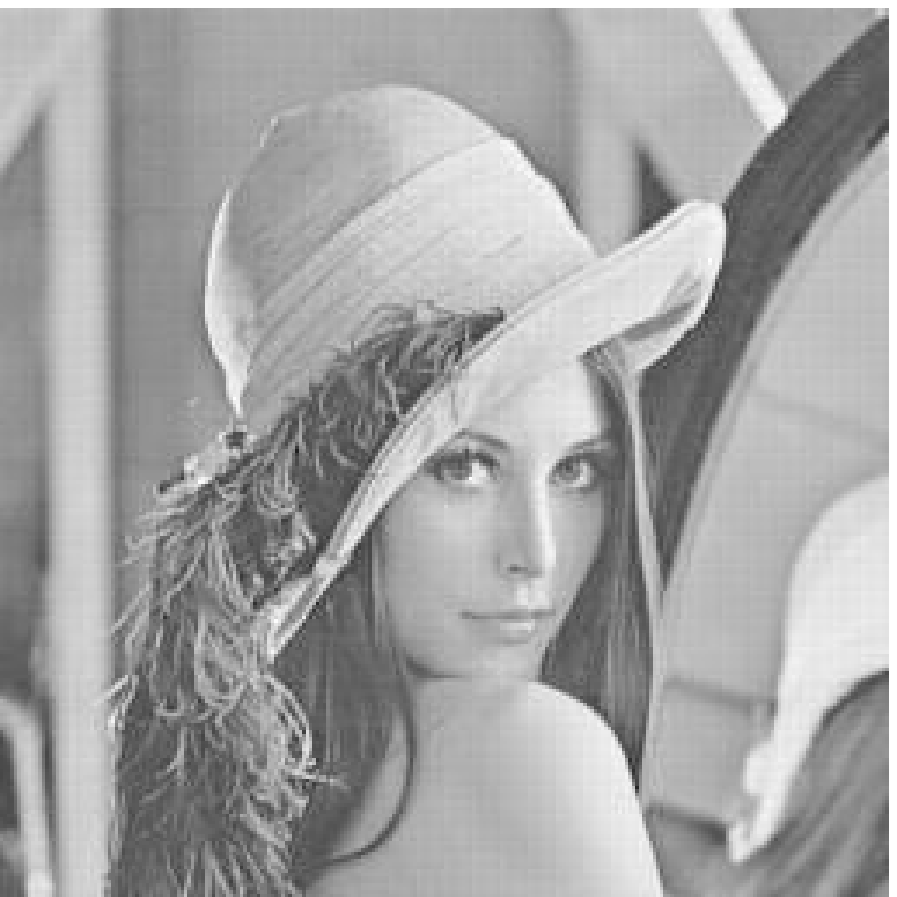}}
			\subfigure[DCT]
	{\label{f:lenaDCT}
		\includegraphics[width=4cm]
		{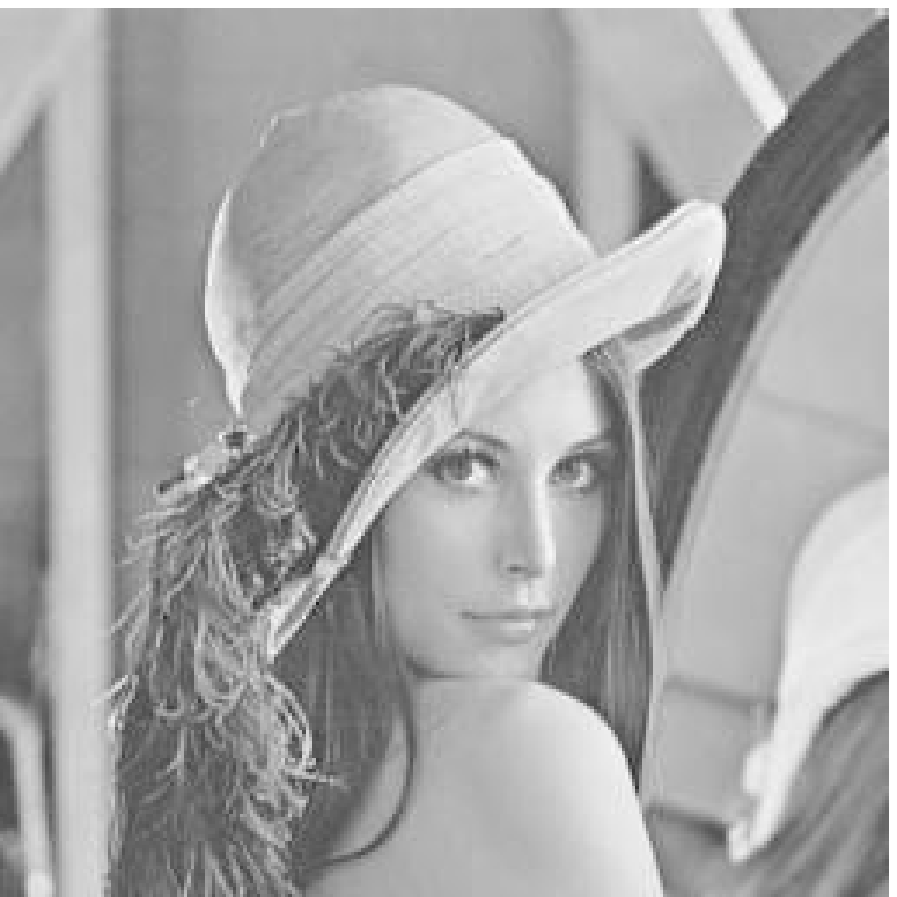}}
	\caption{Compressed \textit{Lena} images.}\label{f:LenasN}
\end{figure*}

\begin{figure*}[h!]
	\centering
	\subfigure[$\widehat{\mathbf{T}}_1$]{\label{f:baboonK1}
		\includegraphics[width=4cm]
		{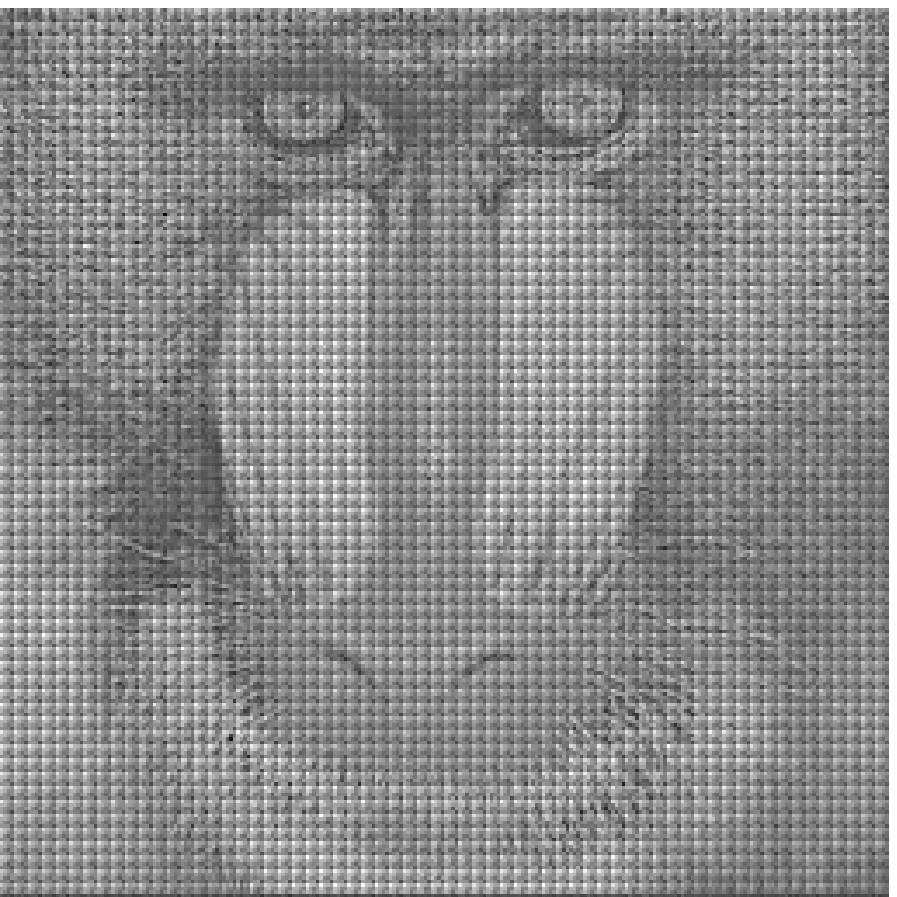}}
	\subfigure[$\widehat{\mathbf{T}}_2$]
	{\label{f:baboonK2}
		\includegraphics[width=4cm]
		{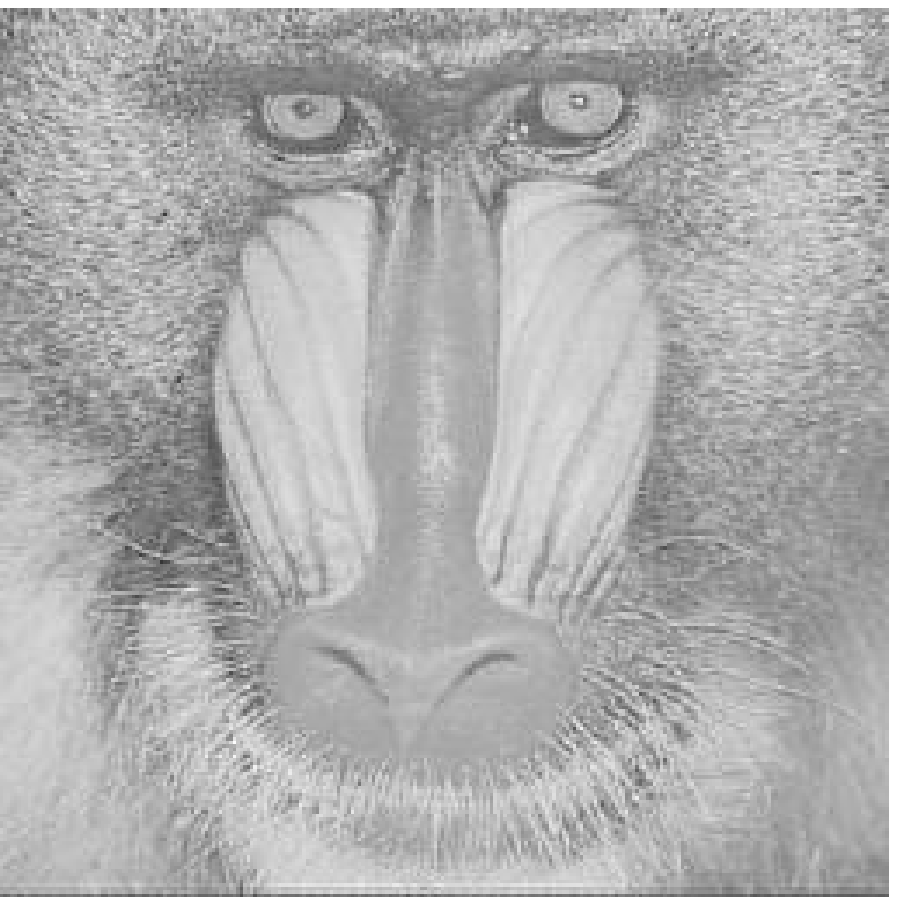}}
	\subfigure[$\widehat{\mathbf{T}}_3$]
	{\label{f:baboonK3}
		\includegraphics[width=4cm]
		{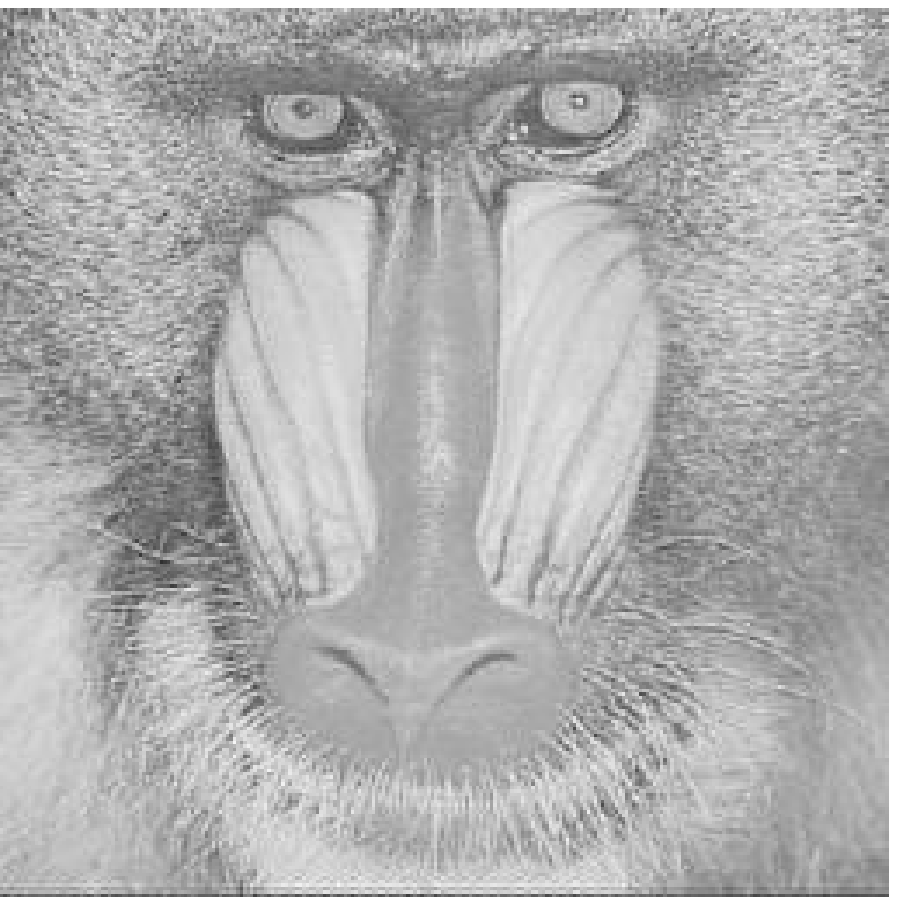}}
	\subfigure[$\widehat{\mathbf{T}}_4$~\cite{cintra2011dct}]{\label{f:baboonK4}
		\includegraphics[width=4cm]
		{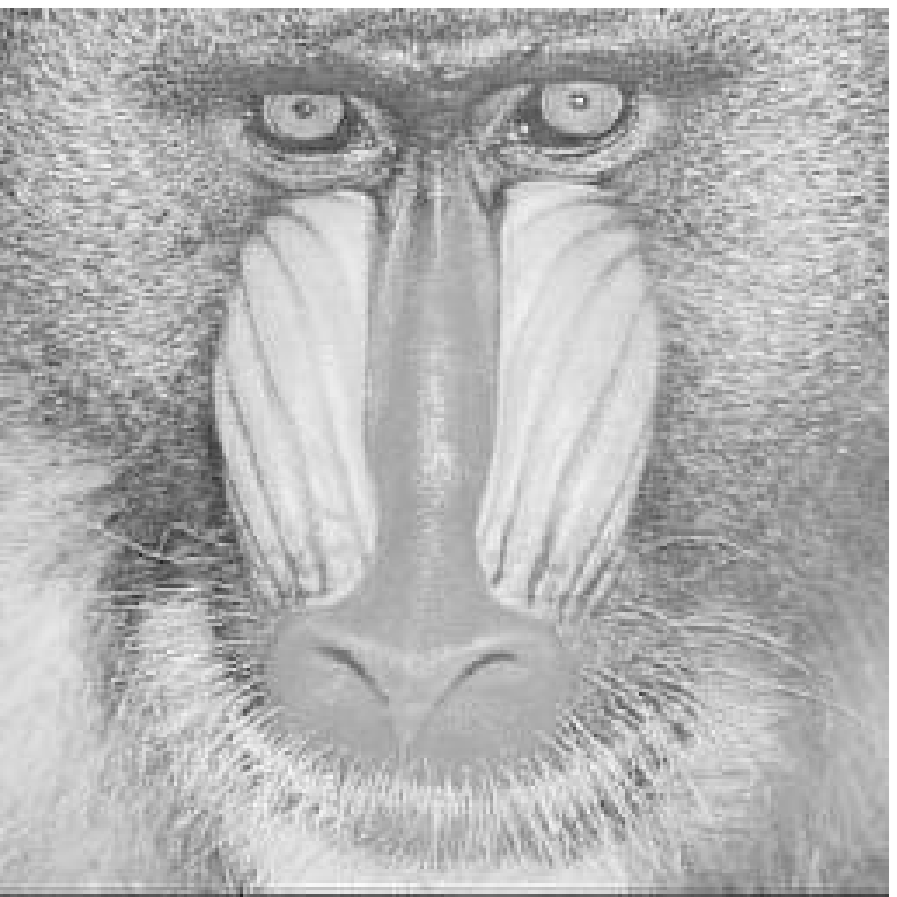}}  \\
	\subfigure[$\mathbf{K}^{(0.3)}$]{\label{f:baboonK03}
		\includegraphics[width=4cm]
		{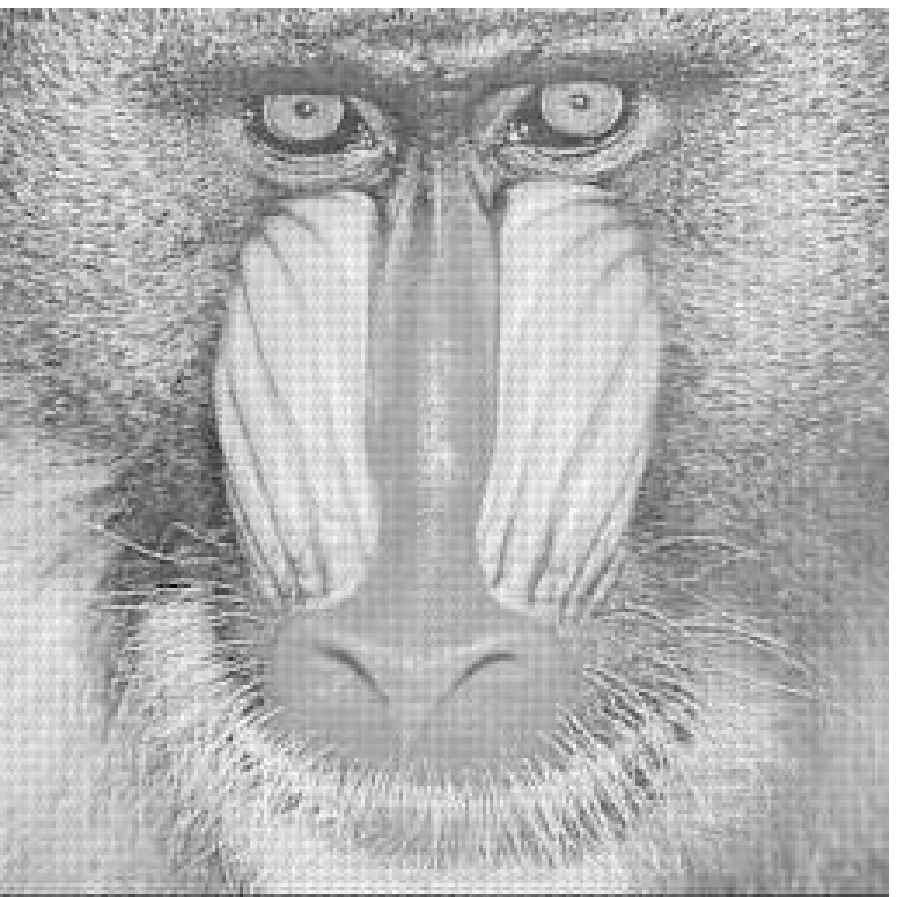}}
	\subfigure[$\mathbf{K}^{(0.4)}$]{\label{f:baboonK04}
		\includegraphics[width=4cm]
		{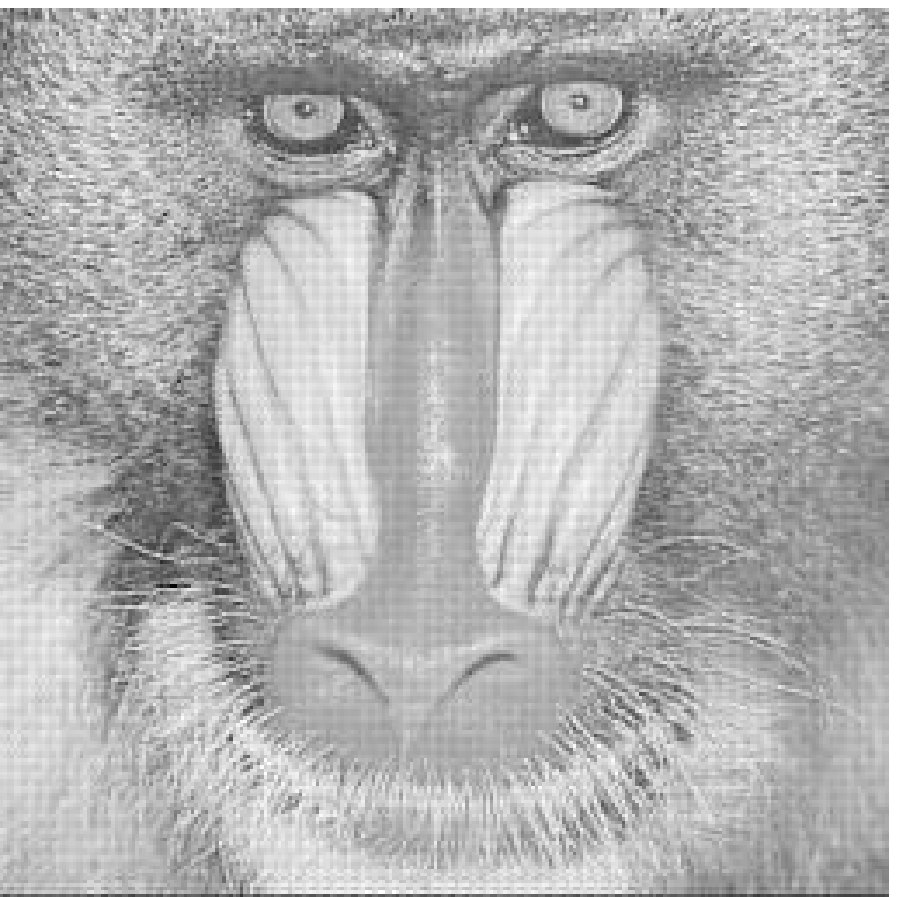}}
	\subfigure[$\mathbf{K}^{(0.7)}$]{\label{f:baboonK07}
		\includegraphics[width=4cm]
		{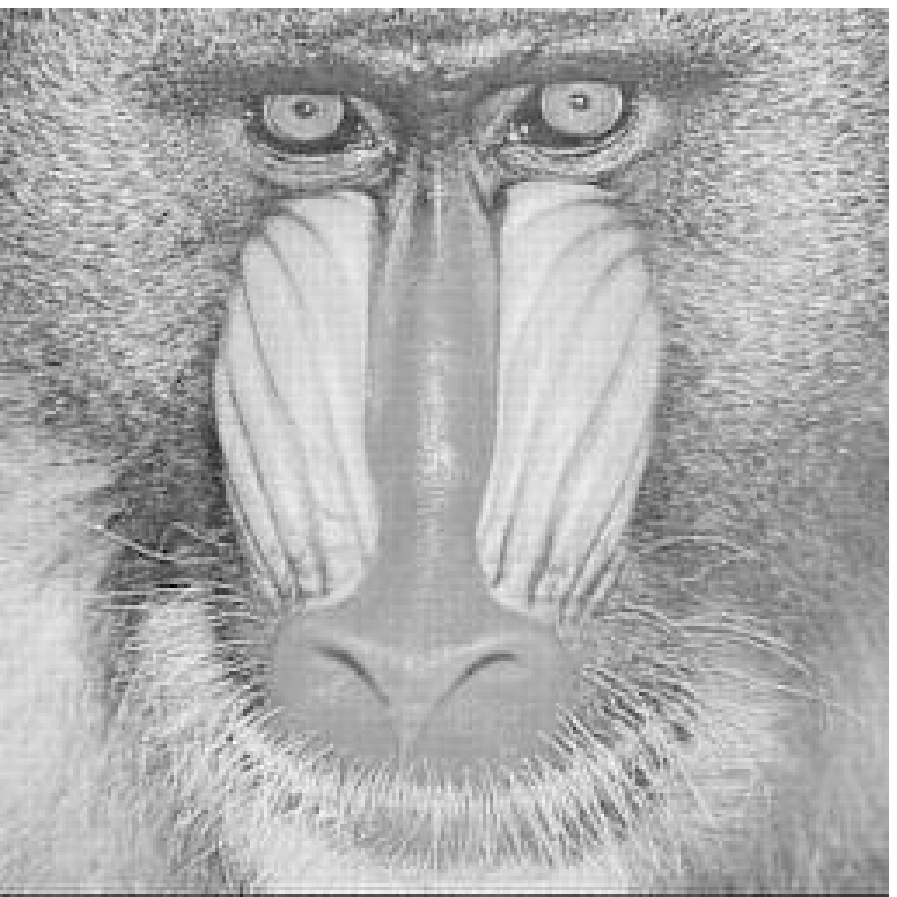}}
	\subfigure[$\mathbf{K}^{(0.8)}$]
	{\label{f:baboonK08}
		\includegraphics[width=4cm]
		{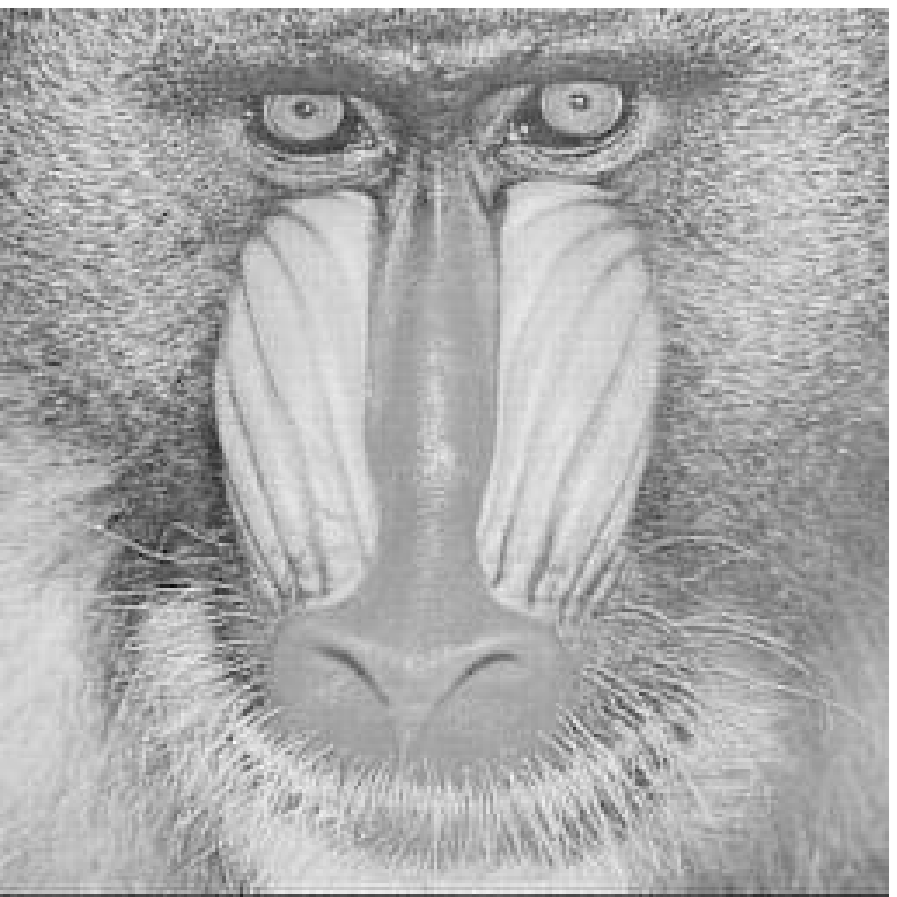}}
	\subfigure[DCT]
	{\label{f:baboonDCT}
		\includegraphics[width=4cm]
		{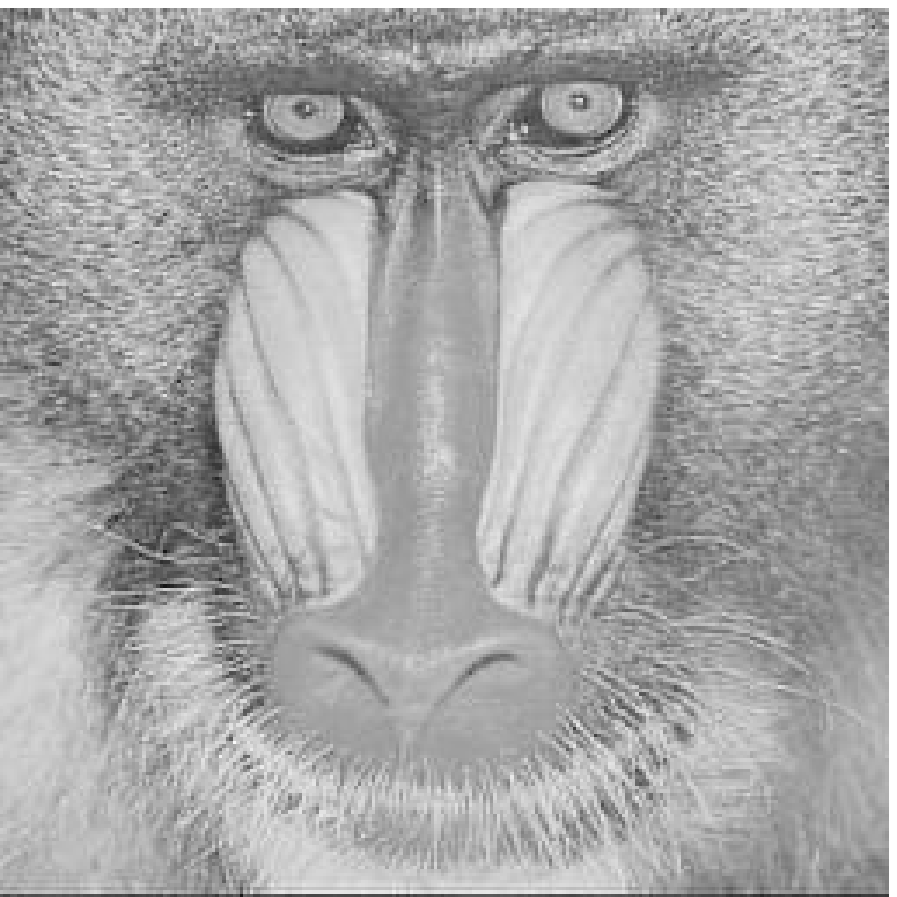}}
	\caption{Compressed \textit{Baboon} images.}\label{f:baboonN}
\end{figure*}

\begin{figure*}[h!]
	\centering
	\subfigure[$\widehat{\mathbf{T}}_1$]{\label{f:moonK1}
		\includegraphics[width=4cm]
		{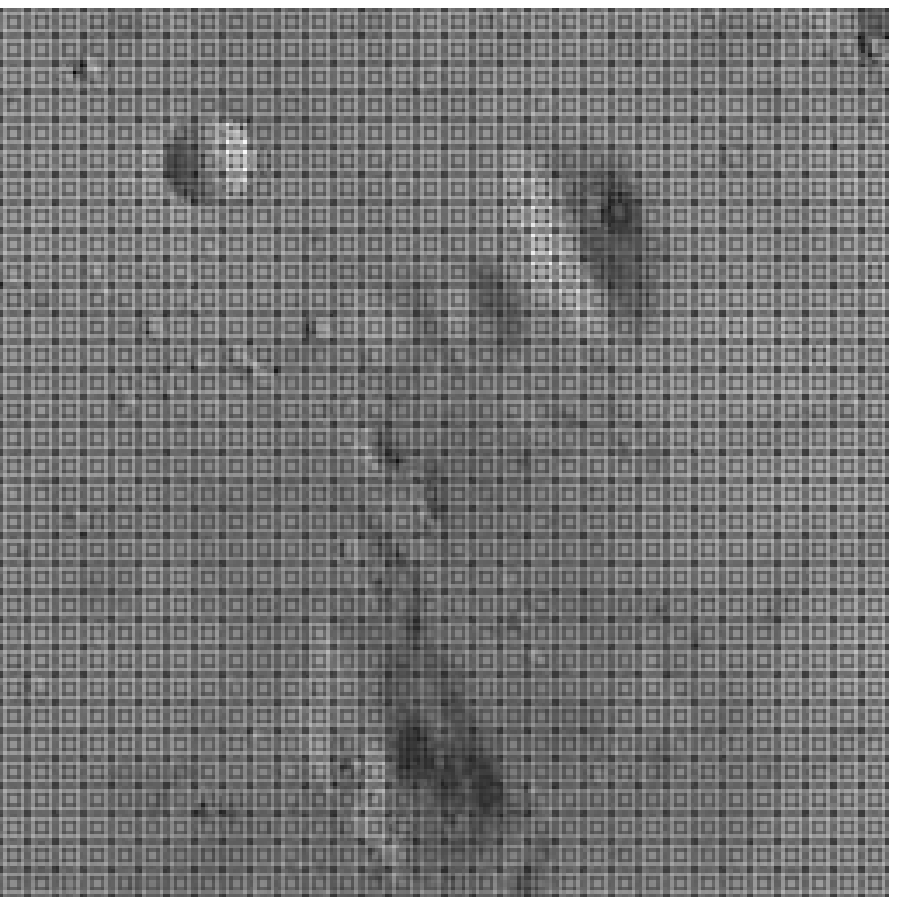}}
		\subfigure[$\widehat{\mathbf{T}}_2$]{\label{f:moonK2}
		\includegraphics[width=4cm]
		{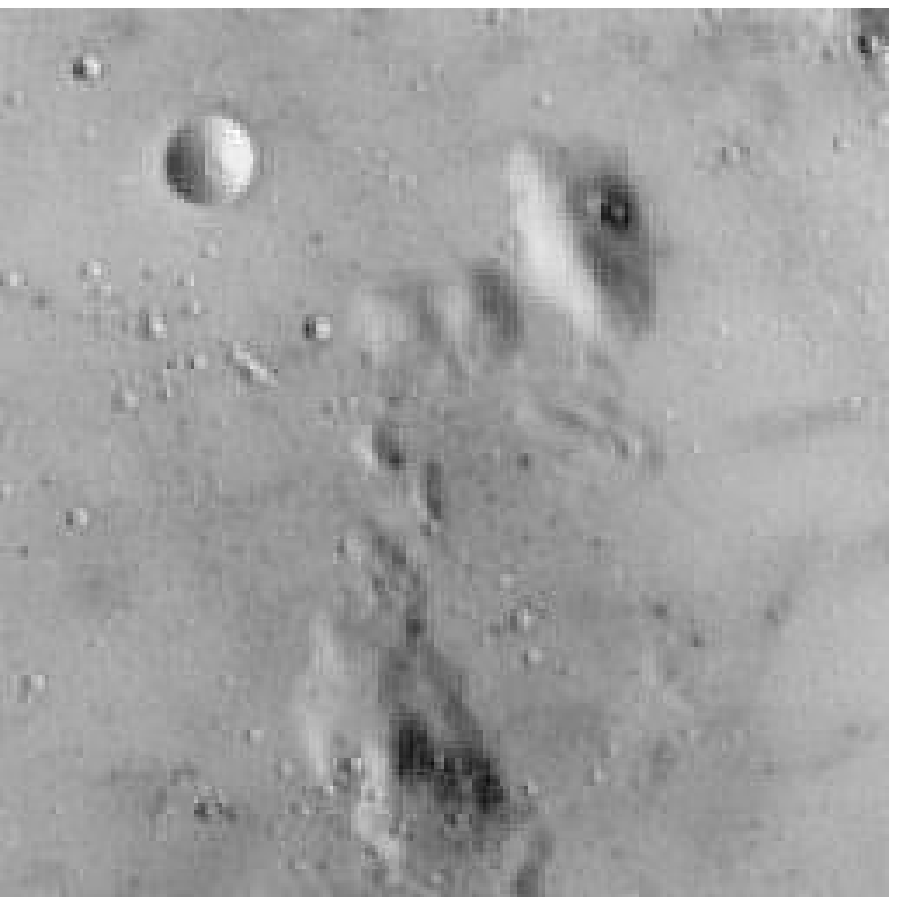}}
		\subfigure[$\widehat{\mathbf{T}}_3$]{\label{f:moonK3}
		\includegraphics[width=4cm]
		{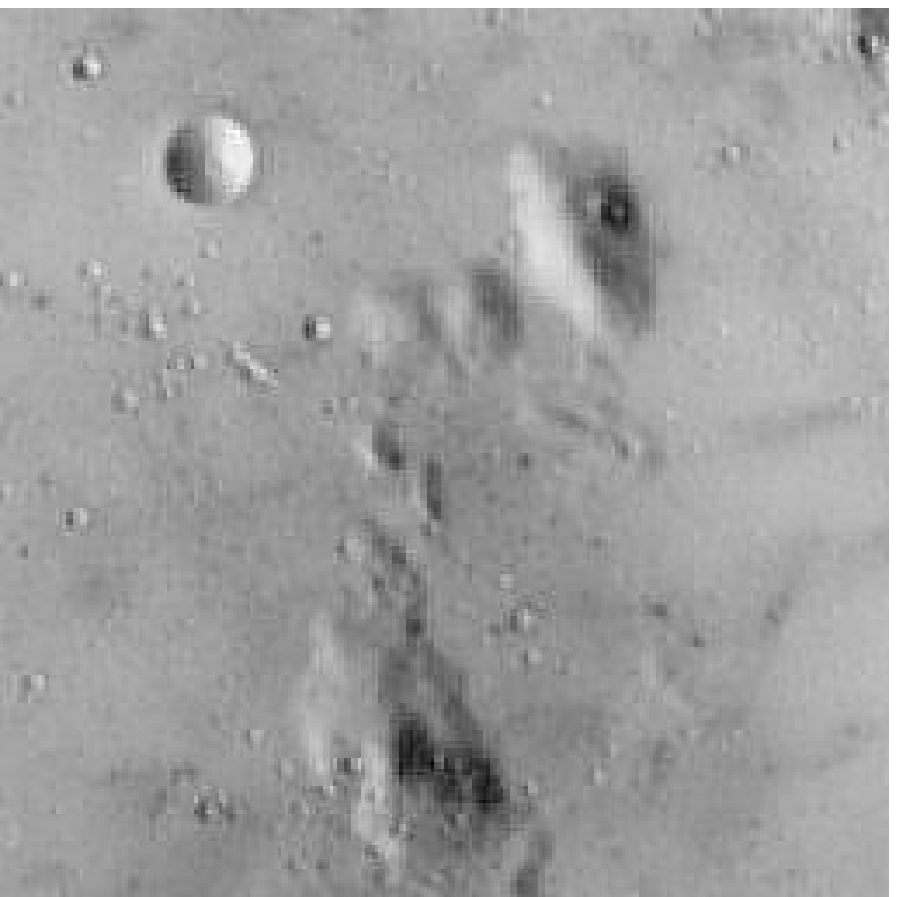}}
		\subfigure[$\widehat{\mathbf{T}}_4$~\cite{cintra2011dct}]{\label{f:moonK4}
		\includegraphics[width=4cm]
		{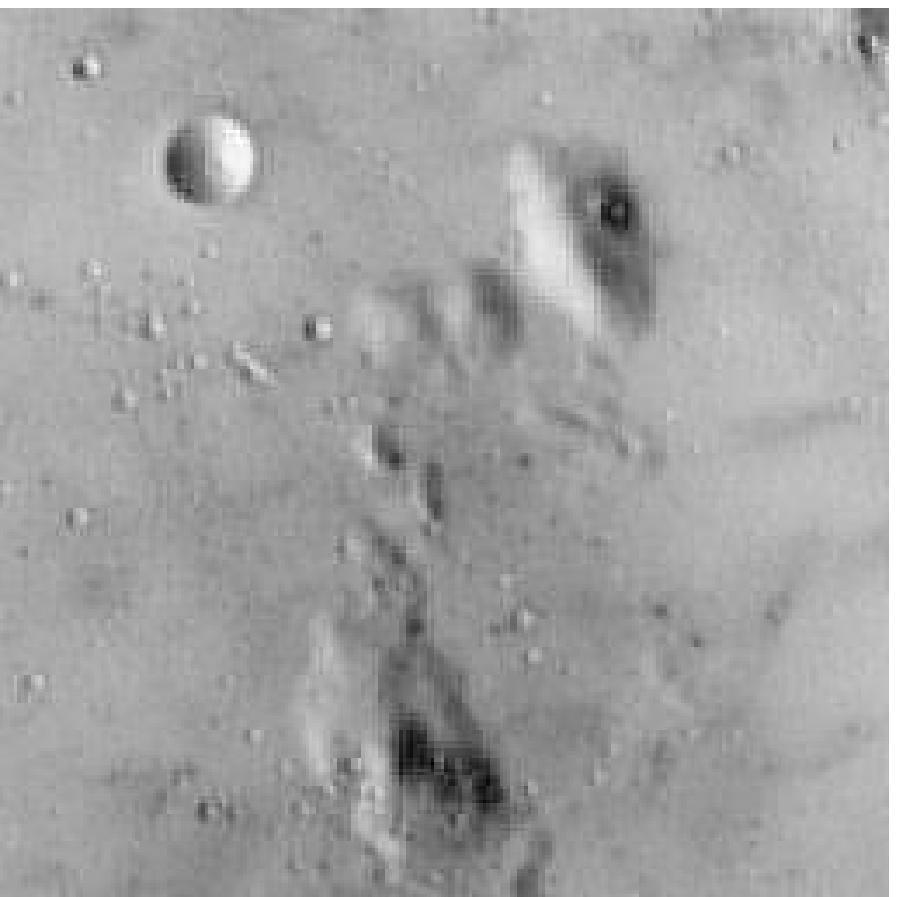}}
 \\
	\subfigure[$\mathbf{K}^{(0.3)}$]{\label{f:moonK03}
		\includegraphics[width=4cm]
		{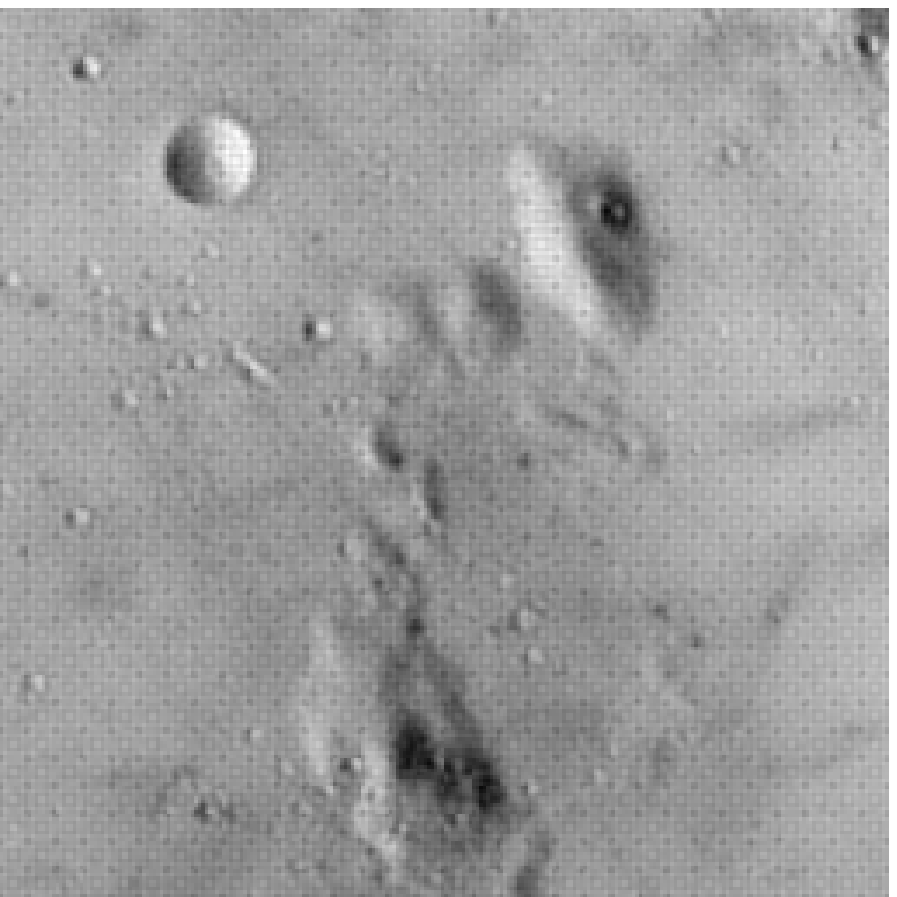}}
		\subfigure[$\mathbf{K}^{(0.4)}$]{\label{f:moonK04}
		\includegraphics[width=4cm]
		{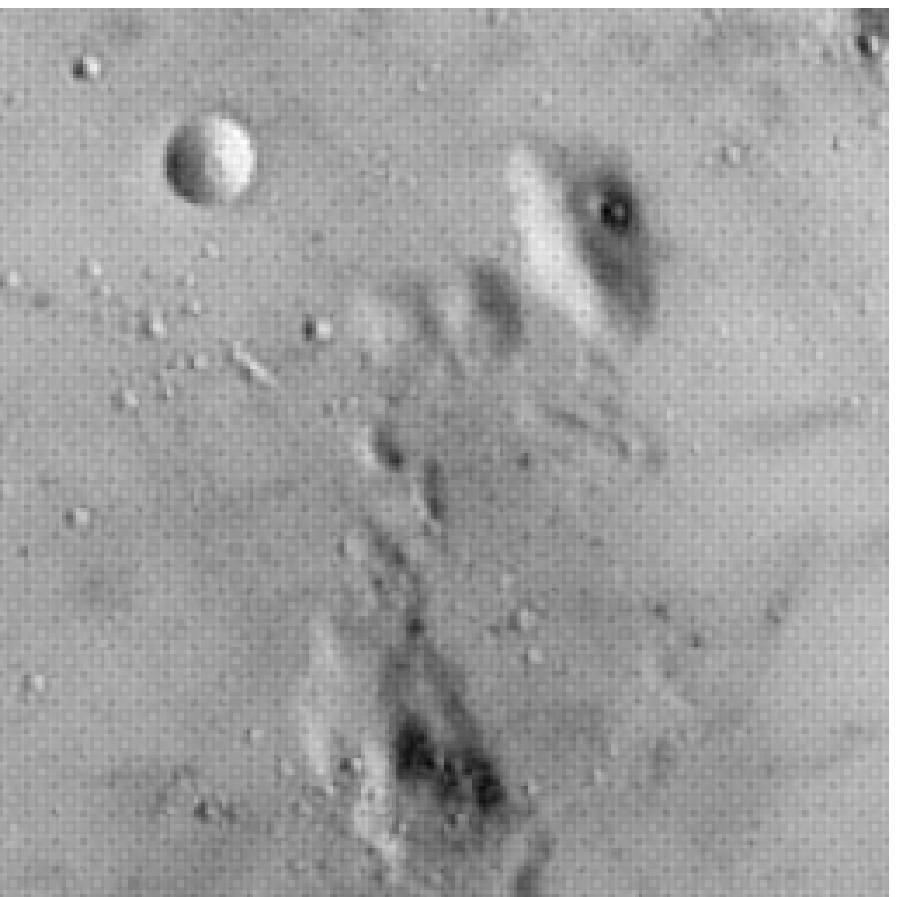}}
		\subfigure[$\mathbf{K}^{(0.7)}$]{\label{f:moonK07}
		\includegraphics[width=4cm]
		{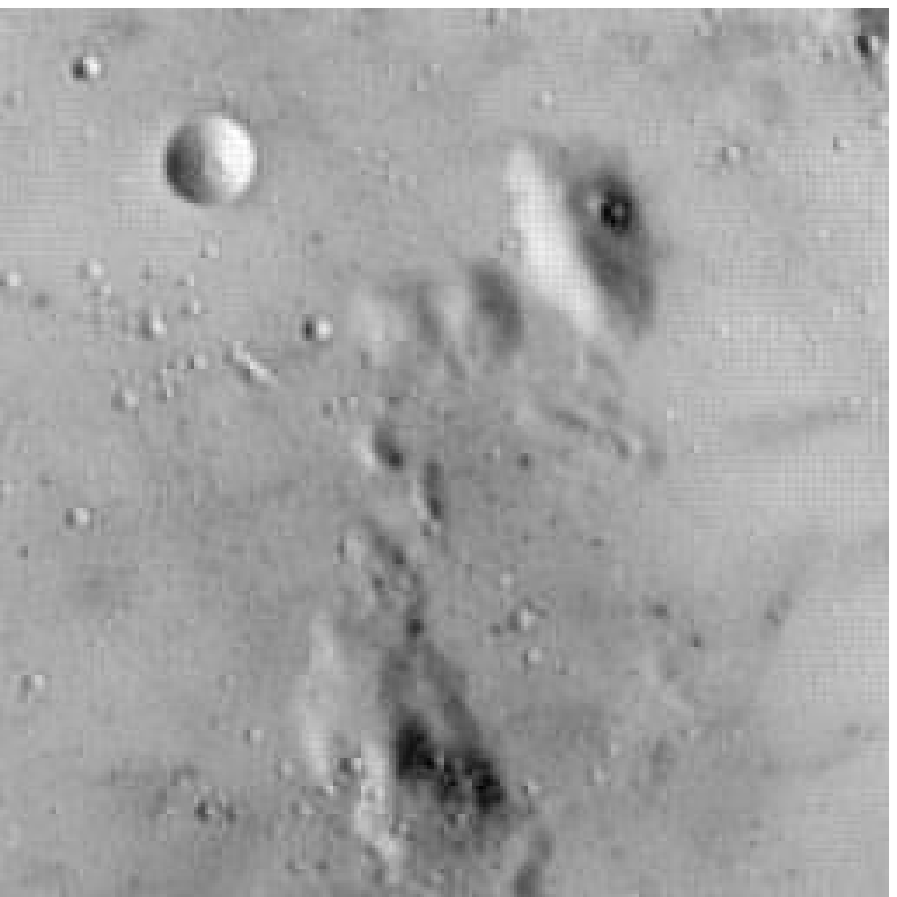}}
		\subfigure[$\mathbf{K}^{(0.8)}$]{\label{f:moonK08}
		\includegraphics[width=4cm]
		{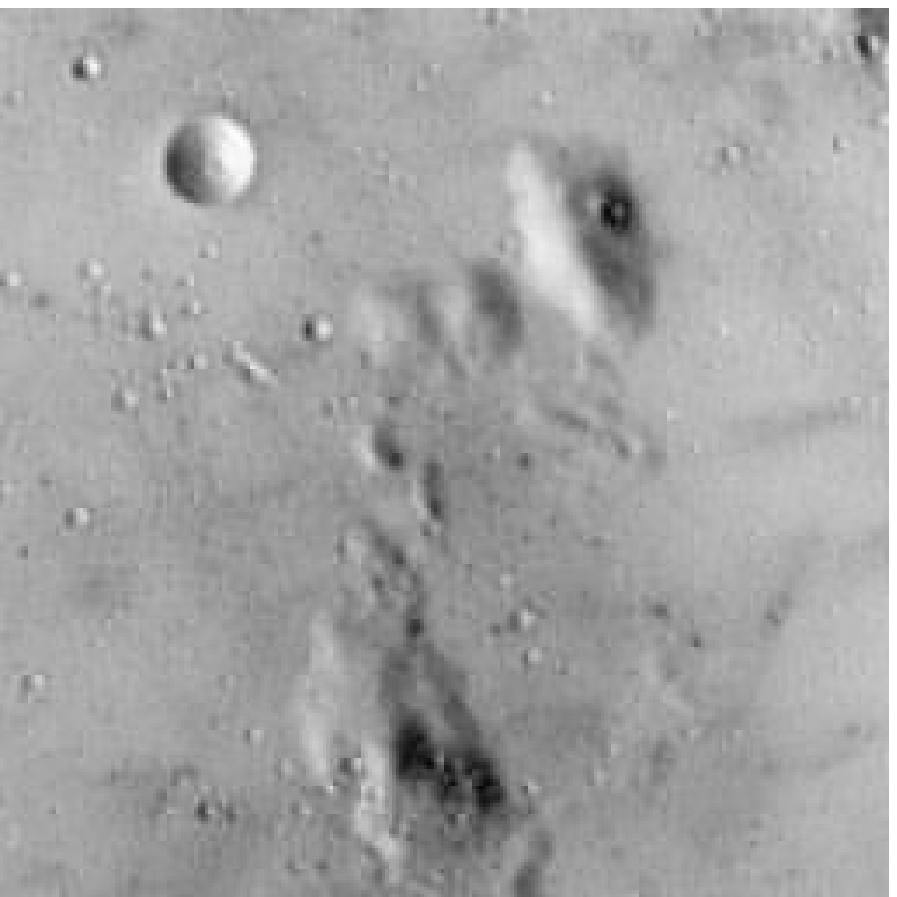}}
		\subfigure[DCT]{\label{f:moonDCT}
		\includegraphics[width=4cm]
		{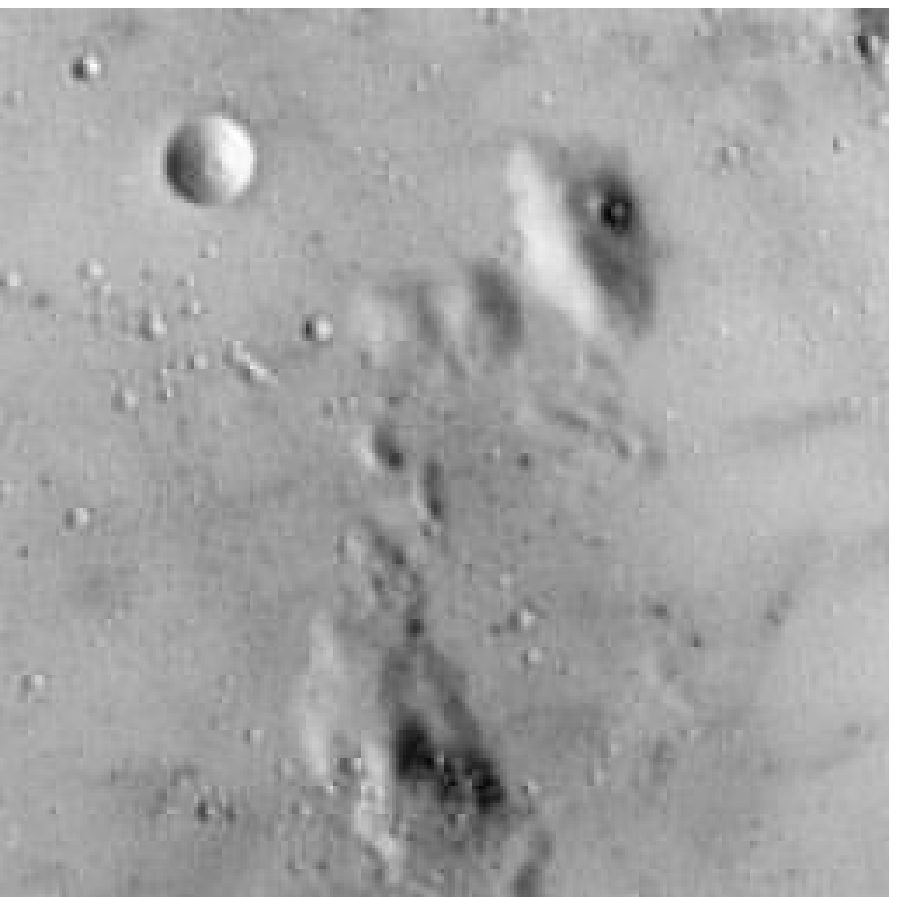}}
	\caption{Compressed \textit{Moon} images.}\label{f:moonN}
\end{figure*}

Fig.~\ref{f:LenasN}, \ref{f:baboonN}, and \ref{f:moonN}
illustrate qualitatively the reconstruction of the
\textit{Lena}, \textit{Baboon}, and \textit{Moon} images, after the application of the compression scheme for $r = 15$ (compression rate of approximately $77 \%$)
using the proposed RKLT, the exact
KLT for $\rho = 0.3$, $0.4$, $0.7$, $0.8$, and the exact DCT.
The assessment metrics from each compressed image are presented in Table~\ref{t:image-measures}.
We highlighted the results from the proposed transforms which performed better than the exact KLT for the value of $\rho$ associated with the interval of which the approximate transform was derived.
Approximations $\widehat{\mathbf{T}}_2$ and $\widehat{\mathbf{T}}_3$ outperformed the exact KLT, $\mathbf{K}^{(0.4)}$ and $\mathbf{K}^{(0.7)}$ res\-pec\-ti\-ve\-ly, according to the values of MSE, PSNR, and MSSIM.
The approximations may outperform the exact KLT because we measure the overall performance of the entire
image compression system,
which includes particular nonlinearities that are better suited for the approximate computation.

\color{black}

\begin{table*}[]
	\centering
	\caption{Quality image assessment measurements for \textit{Lena}, \textit{Baboon}, and \textit{Moon} compressed images}
	\label{t:image-measures}
	\begin{tabular}{llllllllll}
		\toprule
		Image                    & \multicolumn{3}{c}{Lena}                                                       & \multicolumn{3}{c}{Baboon}                                                     & \multicolumn{3}{c}{Moon}                                                       \\ \midrule
		Transform                & \multicolumn{1}{c}{MSE} & \multicolumn{1}{c}{PSNR} & \multicolumn{1}{c}{MSSIM} & \multicolumn{1}{c}{MSE} & \multicolumn{1}{c}{PSNR} & \multicolumn{1}{c}{MSSIM} & \multicolumn{1}{c}{MSE} & \multicolumn{1}{c}{PSNR} & \multicolumn{1}{c}{MSSIM} \\
		\midrule
		$\widehat{\mathbf{T}}_1$ & 3198.18                 & 13.082                   & 0.154                     & 3505.908                & 12.683                  & 0.220                  & 3114.159                & 13.197                  & 0.096                \\
		$\mathbf{K}^{(0.3)}$     & 95.674 		              & 28.323                & 0.660                 & 322.113               & 23.051                 & 0.685                 & 112.060                & 27.636                  & 0.598                 \\
			\addlinespace[1.5ex]
		\midrule
		\addlinespace[1.5ex]
		$\widehat{\mathbf{T}}_2$ & 48.729                  & 31.253                   & \textbf{0.913}                    & 313.852                & 23.164                 & \textbf{0.761}                & \textbf{57.287 }               & \textbf{30.550}                 & \textbf{0.781}                 \\
		$\mathbf{K}^{(0.4)}$     & 70.215                & 29.666                 & 0.720                 & 296.260                & 23.414                 & 0.716                 & 87.816                & 28.695                 & 0.653                 \\
			\addlinespace[1.5ex]
		\midrule
		\addlinespace[1.5ex]
		$\widehat{\mathbf{T}}_3$ & 49.071                & 31.222                 & \textbf{0.913}             & 313.788                 & 23.164                 & 0.761                 & 57.276                & 30.551                 & \textbf{0.781}                 \\
		$\mathbf{K}^{(0.7)}$     & 30.464                & 33.293                 & 0.884                  & 257.115                & 24.030                 & 0.783                 & 50.058                & 31.136                 & 0.776                 \\
			\addlinespace[1.5ex]
		\midrule
		\addlinespace[1.5ex]
		$\widehat{\mathbf{T}}_4$~\cite{cintra2011dct} & 44.593                & 31.638                 & 0.917                 & 286.597                & 23.558                 & 0.766                 & 52.478                & 30.931                 & 0.789                 \\
		$\mathbf{K}^{(0.8)}$     & 25.907                & 33.997                 & 0.916                  & 253.406                & 24.093                 & 0.793                 & 45.705                & 31.531                 & 0.796                 \\
		DCT                      & 23.867                & 34.353                 & 0.938                 & 254.233                & 24.078                 & 0.796                 & 43.543                & 31.742                 & 0.807                 \\
		\bottomrule
	\end{tabular}
\end{table*}

\begin{figure*}[h!]
	\centering
	\subfigure[]{\label{f:MSSIM}\includegraphics[width=7cm]{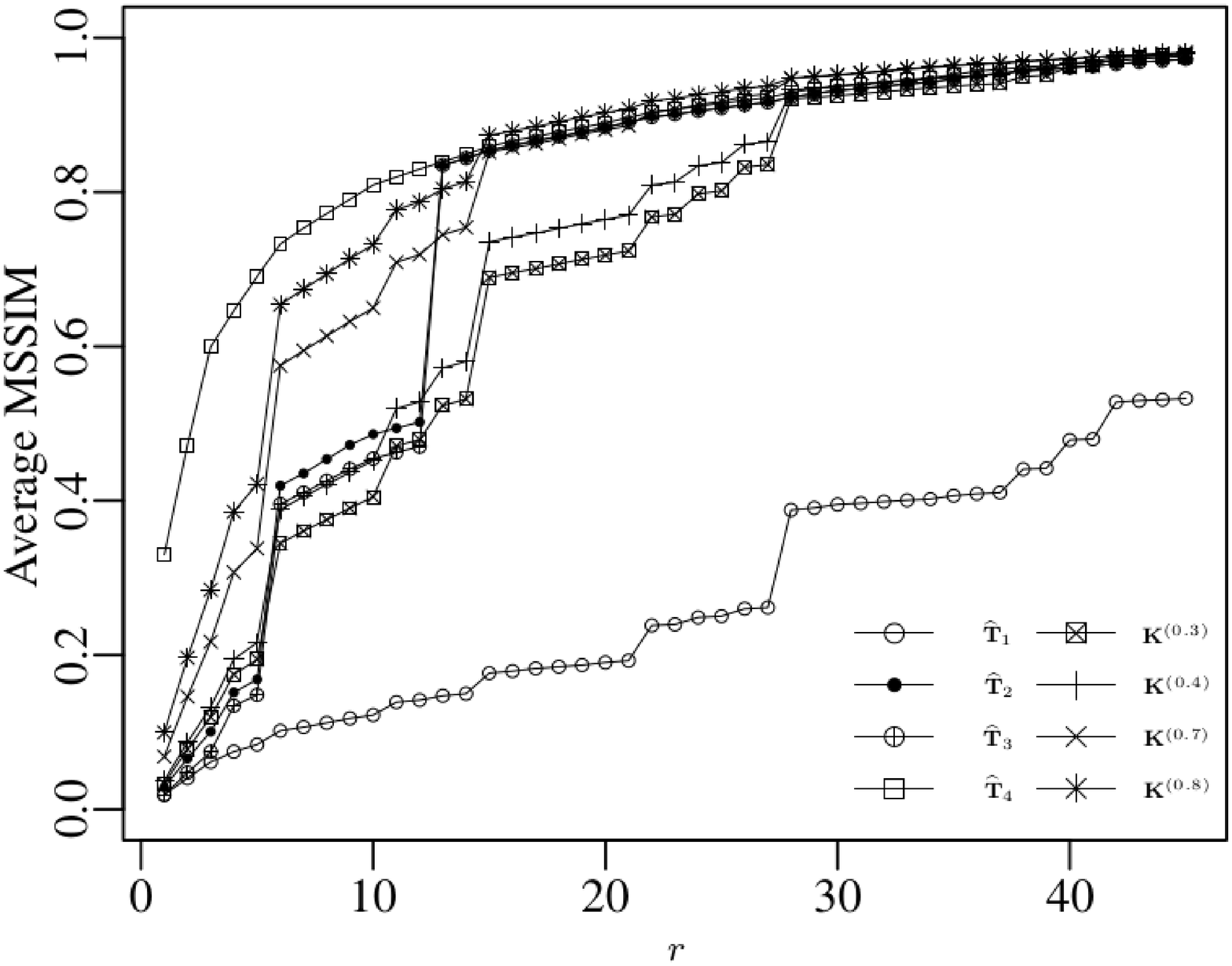}}
	\subfigure[]{\label{f:PSNR}\includegraphics[width=7cm]{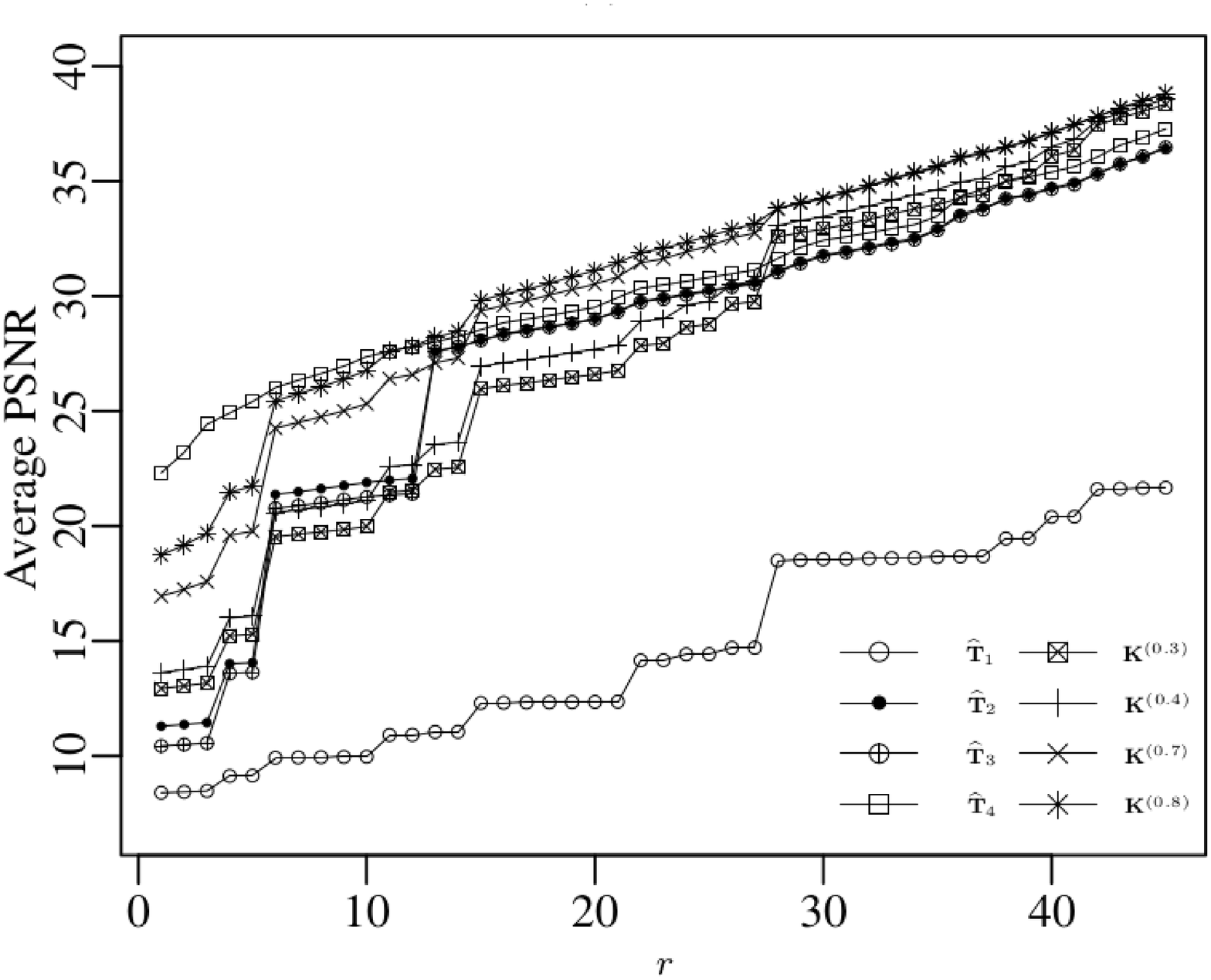}}
	\caption{Image quality measurements for different levels of compression.}
	\label{f:psnrandmssim}
\end{figure*}

For the quantitative analysis, we considered the average
image quality measurements of $45$ compressed standardized images~\cite{uscsipi} considering
different levels of compression ($r \in (0,45)$).
Fig.~\ref{f:psnrandmssim}
presents
the average
image quality measurements
from the compressed images
considering the approximate transforms and
the exact KLT
for values of $\rho = 0.3$, $0.4$, $0.7$, and $0.8$. \color{black}
The approximate transforms
perform similarly to the exact KLT,
mainly when
we
retain
more than $r = 15$ retained coefficients,
except for $\widehat{\mathbf{T}}_1$.
Approximation $\widehat{\mathbf{T}}_4$ outperformed the exact KLT ($\rho = 0.8$) for $r \in [1,11]$ considering PSNR values and for $r \in [1,14]$ considering the MSSIM values.
Considering the performance in JPEG-like compression the proposed approximations exhibited
relevant results, showing a good balance between performance and computational cost.

\section{Conclusions}\label{S:conclusions}

In this paper,
we introduced a methodology
based on the rounding-off function
to design
low-com\-ple\-xi\-ty approximations for the Karhunen-Lo\`eve transform (KLT).
Due to its relevance
in practical image coding systems,
the special case $N = 8$
was comprehensively examined.
According
to qualitative and quantitative
computation experiments,
the proposed transforms were shown
to be good approximations for KLT
as measured by
the adopted quality measures: MSE, PSNR, and MSSIM.
The low-com\-ple\-xi\-ty matrices are natural candidates
for the design of
efficient hardware implementation
capable of operating
at low power consumption
and
high performance.

\section*{Acknowledgements}
We gratefully acknowledge partial financial support from  \textit{Coordena\c c\~ao de Aperfei\c coamento de Pessoal de N\'ivel Superior (CAPES)}, \textit{Conselho Nacional de Desenvolvimento Cient\'ifico e Tecnol\'ogico (CNPq)} and \textit{Funda\-\c c\~ao de Amparo \`a Ci\^encia e Tecnologia de Pernambuco (FA\-CE\-PE)}, Brazil.

\appendix

\section{2D transformation and quantization step}\label{A:appendix}

Let $\mathbf{A}$ be a $8 \times 8$ sub-block from an image.
The 2D transformation from $\mathbf{A}$ induced by an approximation $\widehat{\mathbf{T}}$ is given by:
\begin{align}
	\mathbf{B} &=
	\begin{cases}
		\widehat{\mathbf{T}} \cdot \mathbf{A} \cdot {\widehat{\mathbf{T}}}^\top, &  \text{if $\mathbf{T}$ is orthogonal,}
		\\
		\widehat{\mathbf{T}} \cdot \mathbf{A} \cdot {\widehat{\mathbf{T}}}^{-1}, &   \text{if $\mathbf{T}$ is non-orthogonal,}
	\end{cases} \nonumber
\\
	&=
	\begin{cases}
		(\mathbf{u} \cdot \mathbf{u}^\top) \odot (\mathbf{T} \cdot \mathbf{A} \cdot \mathbf{T}^\top), &  \text{if $\mathbf{T}$ is orthogonal,}
		\\
		(\mathbf{u} \cdot \mathbf{v}^\top) \odot (\mathbf{T} \cdot \mathbf{A} \cdot \mathbf{T}^{-1}),
		&   \text{if $\mathbf{T}$ is non-orthogonal,}
	\end{cases} \nonumber
	\\
&= \mathbf{R} \odot \widehat{\mathbf{B}},  \label{eq:be}
\end{align}
where $\mathbf{u} = \operatorname{diag}(\mathbf{S})$ and $\mathbf{v}$ is given by the inverse elements from $\mathbf{u}$.
In the context of JPEG-like compression~\cite{wallace1992jpeg},
the quantized coefficient matrix $\widehat{\mathbf{B}}$ is given by:
\begin{eqnarray} \label{eq:bbar}
	\bar{\mathbf{B}} = \operatorname{round}(\mathbf{B} \div \mathbf{Q}),
\end{eqnarray}
where $\mathbf{Q}$ is a quantization matrix and $\div$ denotes the element-wise matrix division.

By applying Equation \eqref{eq:be} in \eqref{eq:bbar}, we obtain
\begin{align*}
\bar{\mathbf{B}} = \operatorname{round}(\mathbf{R} \odot \widehat{\mathbf{B}} \div \mathbf{Q}) =
\operatorname{round}(\widehat{\mathbf{B}} \div \tilde{\mathbf{Q}}),
\end{align*}
where $\tilde{\mathbf{Q}} = \mathbf{Q} \div \mathbf{R}$.
Note that $\mathbf{R}$ can be absorbed in the quantization step, thus, the complexity of matrix $\mathbf{S}$ can be dismissed in the image compression applications~\cite{salomon2004data,sayood2017introduction,gonzalez2002digital}.

{\small
\singlespacing
\bibliographystyle{ieeetr}
\bibliography{references-rklt.bib}
}

\end{document}